\documentclass[reprint,superscriptaddress,preprintnumbers,longbibliography,
amsmath,amssymb,aps,floatfix,prl,twocolumn, tightenlines 
]{revtex4-1}
\pdfoutput=1
\usepackage{color}
\usepackage{xcolor}
\usepackage{pdfcolmk}
\usepackage{verbatim}
\usepackage{mathtools}
\usepackage{amsmath}
\usepackage{amssymb}
\usepackage{float}
\usepackage{array}
\usepackage{mathrsfs}
\usepackage{graphicx}
\usepackage{dcolumn}
\usepackage{bm}
\usepackage{amsmath,amsthm}
\usepackage{amssymb,amsfonts}
\usepackage{bbm}
\usepackage{graphicx}
\usepackage{subfigure}
\allowdisplaybreaks[1]
\usepackage[mathcal]{euscript}

\newtheorem*{ansatz*}{Ansatz}

\newcommand{\be}{\begin{equation}}
\newcommand{\ee}{\end{equation}}

\newcommand{\bse}{\begin{subequations}}
\newcommand{\ese}{\end{subequations}}
\newcommand{\ket}[1]{\left|{#1}\right\rangle}

\newcommand{\bra}[1]{\langle{#1}|}

\newcommand{\bpm}{\begin{pmatrix}}
\newcommand{\epm}{\end{pmatrix}}
\newcommand{\bmm}{\begin{matrix}}
\newcommand{\emm}{\end{matrix}}

\makeatletter

\providecolor{lyxadded}{rgb}{0,0,1}
\providecolor{lyxdeleted}{rgb}{1,0,0}

\DeclareRobustCommand{\lyxsout}[1]{\ifx\\#1\else\sout{#1}\fi}

\usepackage[colorlinks,citecolor=blue,linkcolor=blue]{hyperref}

\begin{document}

\title{Non-Abelian Statistics for Bosonic Symmetry-Protected Topological Phases}

\author{Hong-Yu Wang}
\thanks{These authors contribute equally to the work.}
\affiliation{International Center for Quantum Materials and School of Physics, Peking University, Beijing 100871, China}
\affiliation{International Quantum Academy, Shenzhen 518048, China}
\affiliation{Hefei National Laboratory, Hefei 230088, China}
\author{Bao-Zong Wang}
\thanks{These authors contribute equally to the work.}
\affiliation{International Center for Quantum Materials and School of Physics, Peking University, Beijing 100871, China}
\affiliation{Hefei National Laboratory, Hefei 230088, China}
\author{Jian-Song Hong}
\affiliation{International Center for Quantum Materials and School of Physics, Peking University, Beijing 100871, China}
\affiliation{Hefei National Laboratory, Hefei 230088, China}
\author{Xiong-Jun Liu}
\thanks{Corresponding author: xiongjunliu@pku.edu.cn}
\affiliation{International Center for Quantum Materials and School of Physics, Peking University, Beijing 100871, China}
\affiliation{International Quantum Academy, Shenzhen 518048, China}
\affiliation{Hefei National Laboratory, Hefei 230088, China}

\begin{abstract}
Symmetry-protected non-Abelian (SPNA) statistics opens new frontiers in quantum statistics and enriches the schemes for topological quantum computing. In this work, we propose a new paradigm of SPNA statistics in one-dimensional correlated bosonic symmetry-protected topological (SPT) phases and uncover exotic universal features from a systematic investigation. In particular, we show that for generic bosonic SPT phases described by real Hamiltonians, the SPNA statistics of topological zero modes fall into two distinct classes. The first class exhibits conventional braiding of hard-core bosonic zero modes. Furthermore, we discover a second class of unconventional braiding statistics characterized by a nonlinear transformation, featuring a fractionalization of the first class and reminiscent of the non-Abelian statistics of symmetry-protected Majorana pairs. The two distinct classes of statistics have topological origin in classifying non-Abelian Berry phases for braiding processes of real-Hamiltonian systems, distinguished by whether the holonomy involves a reflection operation. To illustrate, we focus on a specific bosonic SPT phase with particle-hole symmetry, and demonstrate that both classes of braiding statistics can be feasibly realized in a tri-junction with and without the aid of a controlled defect, respectively.
Analytic and numerical results are given. We demonstrate how to encode logical qubits and implement both single- and two-qubit gates using the two classes of SPNA statistics. Finally, we propose feasible experimental schemes to observe these predictions and identify the parameter regimes for the high-fidelity braiding, paving the way for the experimental observation of our results in the near future.
\end{abstract}

\maketitle

\section{Introduction}\label{sec:intro}
The pursuit of non-Abelian statistics has been a central theme in quantum physics for decades, driven by both theoretical interest in fundamental physics and potential applications in topological quantum computing~\cite{Wilczek1990,Moore1991362,nayak19962n,ivanov2001non,Kitaev20032,sarma2005topologically,nayak2008non,alicea2011non}. Non-Abelian statistics revolutionizes the traditional particle classification and defines the concept of non-Abelian anyons.
Among the candidates for non-Abelian anyons, Majorana zero modes (MZMs) in topological superconductors have been highly anticipated~\cite{Read2000,kitaev2001unpaired,ivanov2001non,alicea2011non,alicea2012new,Sarma2015,aasen2016milestones,Sato2016}. Despite extensive experimental efforts~\cite{mourik2012signatures,deng2012anomalous,rokhinson2012fractional,das2012zero,wang2012coexistence,churchill2013superconductor,xu2014artificial,nadj2014observation,chang2015hard,albrecht2016exponential,wiedenmann20164,bocquillon2017gapless,zhang2018observation,wang2018evidence,fornieri2019evidence,ren2019topological,jack2019observation,aghaee2023inas}, the experimental signatures of MZMs remain illusive. Key challenges include their susceptibility to disorder~\cite{yu2021non,sarma2021disorder,ahn2021estimating,pan2021disorder} and the soft-gap problem from the proximity effect~\cite{takei2013soft,stanescu2014soft,liu2017phenomenology}. Meanwhile, although digital simulations of non-Abelian braiding have been actively studied on quantum computing platforms recently~\cite{google2023non,xu2023digital,iqbal2024non}, only analog simulations can implement non-Abelian quasiparticles and their braiding in quantum many-body phases with gap protection, which remains a challenge in experiment. These difficulties have motivated the search for alternative approaches.

Recent studies on Majorana Kramers pairs (MKPs) in time-reversal invariant topological superconductors~\cite{Qi2009,Schnyder2010,Wong2012,Zhang2013,Keselman2013,liu2014non,Klinovaja2014,Arbel2019} introduce a novel notion of quantum statistics known as symmetry-protected non-Abelian (SPNA) statistics~\cite{liu2014non}, which has since been widely studied~\cite{liu2014non,gao2016symmetry,hong2022unitary,Haim2019Time,Masaki2024}. The theory of SPNA statistics reveals that in many topological phases, quasiparticles exhibit non-Abelian statistics only under symmetry protection, emphasizing the crucial role of symmetries in quantum statistics. This opens new possibilities for discovering non-Abelian anyons and advancing the applications of symmetry-protected MZMs in quantum computing~\cite{Lapa2021Symmetry,Tanaka2022Manipulation,Schrade2022Quantum}. There are two different categories of SPNA statistics,  protected by unitary symmetries and by anti-unitary symmetries, respectively. The non-Abelian statistics of MKPs is a prototypical example of the latter, as protected by the time-reversal symmetry. The braiding operation essentially corresponds to a dynamical evolution, for which the braiding of MKPs requires special care. Although time-reversal symmetry protects MKPs at each instantaneous step, the complete time evolution is characterized by a unitary operator which generally does not commute with the anti-unitary time-reversal transformation~\cite{gao2016symmetry}. This is known as dynamical symmetry breaking~\cite{gao2016symmetry,mcginley2018topology}, and leads to local mixing in MKPs~\cite{wolms2014local,wolms2016braiding,knapp2020fragility}, for which the SPNA statistics of this category necessitates additional conditions to avoid dynamical symmetry breaking~\cite{gao2016symmetry}. By contrast, the SPNA statistics in unitary symmetric systems do not suffer dynamical symmetry breaking in braiding, hence such Majorana pairs or multiplets are intrinsic SPNA anyons without additional symmetry requirements~\cite{hong2022unitary}. The SPNA statistics underpins the essential mechanism of the braiding statistics of Dirac fermion zero modes in topological insulators~\cite{hong2022unitary,klinovaja2013fractional,wu2020double,wu2020non,wu2022non,huang2024nonabelian}. In comparison with the MZMs, the non-Abelian braiding of Dirac fermion zero modes necessitates not only symmetry protection but also the fine-tuning of the Fermi energy to match the Dirac mode energy levels. Experimental realization of such modes remains challenging.
More recently, SPNA statistics has been further extended to strongly correlated fractional topological phases, which host parafermion zero modes (PZMs)~\cite{Hong2024}. Unlike MZMs in topological superconductors, the symmetry-protected braiding statistics of PZMs exhibits fractionalization, representing a new category of fractionalized SPNA statistics. This study implies that SPNA statistics is a broad concept applicable to strongly correlated topological phases and inspires further exploration of new candidates.

In this work, we propose to study the braiding statistics in one-dimensional (1D) bosonic symmetry-protected topological (SPT) phases, and uncover a new type of fractionalized SPNA statistics with unitary symmetry protection. Unlike free-fermion systems, bosonic SPT phases are inherently strongly correlated quantum many-body systems. Their classification is determined by many-body states that remain invariant under symmetry-preserved continuous deformations, and is mathematically characterized by the group cohomology~\cite{Wen2012}. Bosonic SPT phases can host topological zero modes at boundaries, contributing to ground-state degeneracy. The SPNA statistics of such zero modes 
has not been previously studied and is predicted in this work. Recent advancements in quantum simulators enable the accurate implementation of topological zero modes, overcoming challenges such as disorder, proximity effects, and fine-tuning~\cite{de2019observation,cai2019observation,kiczynski2022engineering}. Adiabatic parameter modulation in a tri-junction configuration enables the braiding of zero modes~\cite{alicea2011non}. Consequently, our prediction of SPNA statistics for topological zero modes in bosonic SPT phases holds significant feasibility for experimental realization.

Our core results are as follows. We predict that the topological zero modes in the 1D bosonic SPT phase exhibit two fundamental classes of SPNA statistics -- a universal feature of the bosonic SPT phases with real Hamiltonians and in sharp contrast to the free-fermion topological phases. The first class is the conventional braiding statistics, similar to the one of hard-core bosons. Remarkably, we discover a second class of unconventional braiding statistics, which is characterized by an exotic nonlinear transformation, featuring a fractionalization of the former class and a reminiscent of braiding two Majorana pairs. 
The conventional and unconventional braiding statistics belong to distinct topological classes with different nontrivial non-Abelian Berry phases,
with which the non-commuting braiding matrices can be implemented on a single qubit given by two bosonic zero modes. The predicted exotic braiding statistics unveil an intrinsic dynamical feature of bosonic SPT phases, providing new insights into their characterization and fundamental nature. Further, we consider a topological spin-exchange model, which can be mapped to the hard-core bosonic Su-Schrieffer-Heeger (SSH) model~\cite{Su1979}, to illustrate the general theory. 
In particular, the two classes of SPNA statistics are shown to be achievable by braiding the zero modes via a tri-junction with and without the assistance of a local defect put in the junction point, respectively. This provides a highly feasible way toward the experimental realization of SPNA statistics. Unlike a single chain model, the model in a tri-junction geometry is generically not exactly solvable due to strong correlation, and our analytical predictions are confirmed with numerical results. As expected, we show that the braiding statistics is protected and robust against local symmetry preserving disorders. 
Additionally, the presence of two classes of non-Abelian statistics facilitates the implementation of the single- and two-qubit topological gates. Finally, we propose real experimental setting to observe the SPNA statistics in the bosonic SPT phase, paving a way for experimental realization 
in the near future.

The remainder of this paper is organized as follows. Section~\ref{sec:zeromodes} introduces the topological zero modes in the bosonic SPT phase of a topological spin-exchange model, with the symmetry protection of the topological phase being studied in both static and dynamical regimes. In Sec.~\ref{sec:brazm}, we develop a generic effective Hamiltonian theory for the SPNA statistics
 of the bosonic zero modes, protected by particle number conservation and particle-hole symmetry which are unitary, with the universal feature of two classes of SPNA statistics being obtained. Section~\ref{sec:num} presents a numerical verification of our prediction. The effects of dynamical symmetry breaking and the mechanism of symmetry protection for SPNA statistics are discussed in more detail in Sec.~\ref{sec:dybrking}. Section~\ref{sec:experm}  proposes experimental schemes for realizing the predicted SPNA statistics and highlights their high feasibility for analog quantum simulations. Finally, the conclusion and outlook for future issues are given in Sec.~\ref{sec:outlook}. 

\begin{figure*}
    \centering
    \includegraphics[scale=0.23]{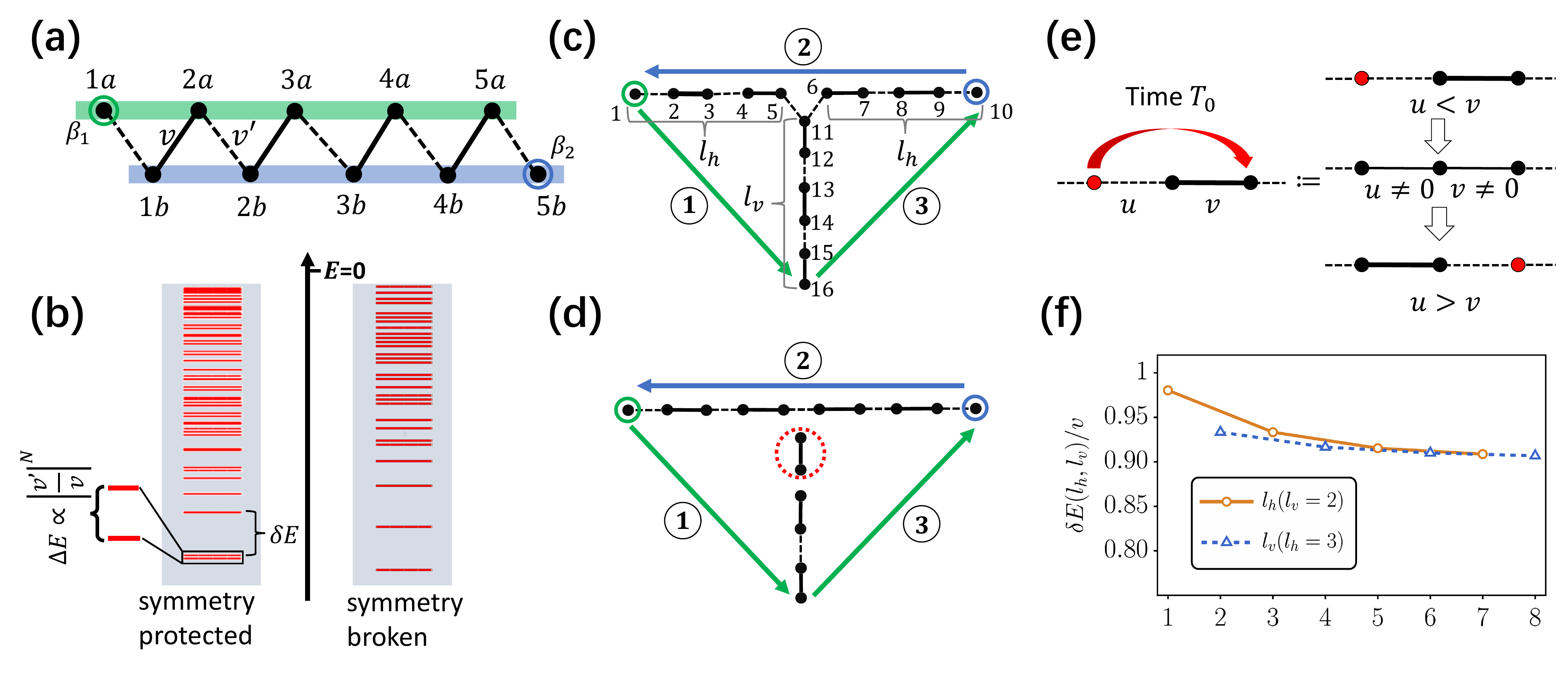}
    \caption{ Illustration of the topological spin-exchange model and braiding schemes. Solid line between sites represents the strong spin-exchange coupling $v>0$, while the dashed line denote weak couplings $v'>0$ with $v' < v$. (a) The topological spin-exchange model ($N=5$) with two sublattices indicated by green and blue strips. The green and blue circles represent the left and right zero modes, respectively, localized at the ends of the system. (b) Low-energy spectrum at half filling (i.e., half of the spins are in the spin-up state) with symmetry protection (left) and symmetry breaking (right). The spectrum is ploted with coupling strengths set to $v=1$ and $v'=0.5$. The half-filled model with symmetry protection exhibit two-fold ground-state degeneracy with an exponentially decaying energy difference $\Delta E$ between the two ground states and a finite bulk gap $\delta E$ between the ground state and the first excited state. (c) The first type of braiding scheme realizes the conventional braiding of zero modes. Green and blue arrows indicate the motion orders and directions of zero modes for braiding. The integers $l_v$ and $l_h$ represent the number of sites in the vertical direction and the half number of sites in the horizontal direction of the tri-junction (with only left-right symmetric tri-junctions considered). The whole system is half-filled. (d) The second type of braiding scheme with a local defect (encircled by the red dotted line) realizes the exotic unconventional braiding. The local defect is empty, while the rest of the tri-junction is half-filled. (e) Schematic of zero mode movement. The red site marks the position of the localized zero mode, which shifts by two sites when coupling strengths $u$ and $v$ are adiabatically tuned. The time $T_0$ represents the duration of the movement. (f) The energy gap $\delta E$ (in units of $v$) between ground states and excited states of the tri-junction in the first braiding scheme as a function of the vertical ($l_v$) and horizontal ($l_h$) sizes. The orange line corresponds to varying $l_h$ with fixed $l_v=2$, while the blue dashed line corresponds to varying $l_v$ with fixed $l_h=3$. The horizontal axis denotes $l_h$ for the orange line and $l_v$ for the blue dashed line.}
    \label{fig:specandbraiding}
\end{figure*}

\section{Topological bosonic zero modes with symmetry protection}\label{sec:zeromodes}

We start in this section to investigate the symmetry protection mechanism of topological zero modes and braiding statistics in the bosonic SPT phase in both the static and dynamical regimes. For clarity, we focus on a spin-exchange model with alternating coupling strengths, whose symmetries include unitary particle-hole symmetry, particle-number conservation, and an anti-unitary symmetry. Although both unitary and antiunitary symmetries can protect the topology of bosonic SPT phases and ensure the existence of zero modes in the static setting, we show that only unitary symmetries can protect the topological phases in the dynamical regime. Since the braiding is inherently dynamical, the SPNA statistics are well defined when the protecting symmetry is preserved throughout the entire braiding operation, both for the static topological phase and for the dynamical braiding evolution, a condition that we show is generically satisfied by unitary symmetries.

\subsection{Bosonic zero modes in the topological spin-exchange model}
We consider a topological spin-exchange model with alternating couplings on a one-dimensional lattice with $2N$ sites [Fig.~\ref{fig:specandbraiding}(a)]. The sites are divided into two sublattices, labeled $a$ and $b$. The Hamiltonian is given by
\begin{eqnarray}\label{eq:expham}
H_{0}&=&\frac{1}{2}\sum_{j=1}^{N} v' (\sigma^x_{j,a}\sigma^x_{j,b}+\sigma^y_{j,a}\sigma^y_{j,b})\nonumber\\
&&+\frac{1}{2}\sum_{j=1}^{N-1} v (\sigma^x_{j,b}\sigma^x_{j+1,a}+\sigma^y_{j,b}\sigma^y_{j+1,a}),
\end{eqnarray}
where $v'$ and $v$ are non-negative real parameters. This model can be mapped to the hard-core bosonic Su-Schrieffer-Heeger (SSH) model~\cite{Su1979} via Holstein-Primakoff transformation. In this mapping, the spin up (down) state is identified with the bosonic vacuum state $\ket {0}$ (occupied state $\ket {1} = b^{\dagger} \ket {0}$), where the hard-core bosonic operators are defined as $b^{\dagger}_i=\sigma_i^-\equiv(\sigma^x_i-i\sigma^y_i)/2$ and $b_i=\sigma_i^+\equiv(\sigma_i^x+i\sigma_i^y)/2$. The mapped Hamiltonian is then given by
\begin{equation}\label{eq:SSHham}
H_{0}=v' \sum_{j=1}^{N} b_{j,a}^{\dagger}b_{j,b}+ v  \sum_{j=1}^{N-1} b_{j,b}^{\dagger}b_{(j+1),a}+h.c.,
\end{equation}
The hard-core exclusion, $b^2_{i}=0$, satisfied by these bosonic operators, endows the model with strong correlations. As a result, Eq.~\eqref{eq:SSHham} represents a strongly interacting Hamiltonian for bosons.

In the topological phase ($v > v'$), the model hosts two hard-core bosonic zero modes, each localized at one end of the chain. We formally label the creation operators of the left and right zero modes as $\beta_1^{\dagger}$ and $\beta_2^{\dagger}$. The zero modes can be either empty ($n_i=0$) or occupied ($n_i=1$), resulting in a four-fold degenerate ground-state subspace spanned by the states ${\lvert n_1, n_2 \rangle}$.  In the case of $v >> v'$ and the thermodynamic limit $N\gg 1$, the zero-mode operators read:
\begin{eqnarray}\label{eq:zeromodedof}
    \beta_1^{\dagger} &=& \mathcal{N}\left(-b_{1,a}^{\dagger} + \sum_{j=2}^N \left(-\frac{v'}{v}\right)^{j-1} b_{j,a}^{\dagger}\right), \nonumber
    \\
    \beta_2^{\dagger} &=& \mathcal{N}\left(-b_{N,b}^{\dagger} + \sum_{j=1}^{N-1} \left(-\frac{v'} {v}\right)^{N-j} b_{j,b}^{\dagger}\right),
\end{eqnarray}
where $\mathcal{N} \approx \sqrt{1-\left(\frac{v'}{v}\right)^2}$ is the normalization factor. The zero-mode operators obey the same commutation relations as hard-core bosons, commute with $H_0$ within the ground-state manifold, and reduce to $\beta_1^{\dagger} = b^{\dagger}_{1,a}$ and $\beta_2^{\dagger} = b^{\dagger}_{N,b}$ in the dimerized limit. When the system is finite, the finite-size effect hybridizes the left and right zero modes, opening an exponentially small energy splitting $\Delta E\propto\left(\frac{v'}{v}\right)^N$ [left panel of Fig.~\ref{fig:specandbraiding}(b)], which is negligible for typical large system size. The zero end modes are then protected by the bulk topological gap $\delta E$.

\subsection{Symmetry protection of zero modes}
The Hamiltonian Eq.~\eqref{eq:SSHham} preserves particle number conservation and particle-hole symmetry, whose corresponding symmetry operations are denoted by $U_{\theta}$ and $C$, respectively,  and defined as follows:
\begin{eqnarray} \label{eq:systransform}
 U_{\theta} b_i U_{\theta}^{-1} = e^{-i\theta} b_i, & \quad & C b_{i} C^{-1}=b_{i}^{\dagger}.
\end{eqnarray}
Both symmetries are unitary and take the representation:
\begin{equation}
 U_{\theta}=\text{exp}(i\theta \sum_{i}b_i^{\dagger}b_i), \quad  C=\prod_{i}(b_{i}+b_{i}^{\dagger}).
\end{equation}
The two unitary symmetries are sufficient to protect the zero modes. To demonstrate this, we employ the effective-Hamiltonian approach to symmetry analysis: the dynamics of zero modes is governed by symmetry-allowed terms constructed from the zero-mode operators $\beta_1$, $\beta_2$, $\beta_1^{\dagger}$, and $\beta_2^{\dagger}$. The most general terms in the effective Hamiltonian capable of lifting the zero modes take the form
\begin{equation}
H_{j}=m_{j}\beta_{j}+m_{j}^{*}\beta_{j}^{\dagger}+\epsilon_{j}(\beta_{j}^{\dagger}\beta_{j}-1/2),
\end{equation}
where $j=1$, $2$, $m_j$ are arbitrary complex coefficients, and $\epsilon_j$ are arbitrary real parameters, together with the pairing term
\begin{equation}
H_{\Delta}=\Delta \beta_1\beta_2 + \Delta^* \beta_1^{\dagger}\beta_2^{\dagger}.
\end{equation}
Symmetry protection is manifested in the fact that these perturbations do not commute with the symmetry operators.
To illustrate the necessity of both unitary symmetries for protecting the zero modes and the resulting ground-state degeneracy, we numerically compute the low-energy spectrum at half filling in the presence of a symmetry-breaking term,
\begin{equation}
H'= \epsilon \sum_{j,a} b^{\dagger}_{j,a} b_{j,a},
\end{equation}
with $\epsilon=0.5$. This term breaks particle-hole symmetry while preserving particle-number conservation. As shown in the right panel of Fig.~\ref{fig:specandbraiding}(b), the ground-state degeneracy is lifted by this term.

Because the Hamiltonian is real, it is naturally invariant under the complex conjugate operation $K$. Composition of $K$ with particle-hole operation yields the anti-unitary symmetry generator:
\begin{equation}
C^{A} = \prod_i(b^{\dagger}_i + b_i)K.
\end{equation}
Together with particle-number conservation, this antiunitary symmetry can, in principle, protect the zero modes. Nevertheless, such protection is not robust under dynamical evolution owing to dynamical symmetry breaking. To illustrate this point, we consider the time-dependent Hamiltonian:
\begin{eqnarray}
H_{\text{dsb}}(t)=H_{0}&+&\delta_{1}(b_{1,a}^{\dagger}b_{1,b}+b_{1,b}^{\dagger}b_{1,a})\cos\omega t \nonumber
\\
&+&\delta_{2}(ib_{1,a}^{\dagger}b_{1,b}-ib_{1,b}^{\dagger}b_{1,a})\sin\omega t,
\end{eqnarray}
where $\delta_{1,2}$ are real numbers. This Hamiltonian commutes with $C^A$ and $U_{\theta}$ at
arbitrary time $t$ but breaks $C$ for $\sin\omega t\neq0$. When $\omega\gg\delta_{1(2)}$, the effective Hamiltonian obtained by Floquet theory (Methods~\ref{sec:Floquet}) reads
\begin{eqnarray}\label{eq:dbana}
H_{\text{eff}} & = & H_{0}+\frac{\delta_{1}\delta_{2}}{\omega}(b_{1,a}^{\dagger}b_{1,a}-b_{1,b}^{\dagger}b_{1,b})+\mathcal{O}\left(\frac{1}{\omega^{2}}\right).
\end{eqnarray}
The second term in $H_{\text{eff}}$ destroys the zero modes, implying that the anti-unitary symmetry in the static regime is generically broken down in the dynamical regime. In Sec.~\ref{sec:dybrking}, we further study dynamical symmetry breaking caused by complex random hopping disorders.

The role of anti-unitary symmetries in this context is to enable a more refined mathematical classification. Specifically, without the complex conjugation $K$, the symmetry operators $U_{\theta}$ and $C$ do not commute, yielding the symmetry group $U(1)\rtimes \mathbb{Z}_2$ with group cohomology classification $H^2(U(1)\rtimes \mathbb{Z}_2, U(1))=\mathbb{Z}_2$. Including $K$ allows one to choose a representation $\tilde U_{\theta} = \text{exp}(i\theta\sum_i(b^{\dagger}_ib_i-\frac{1}{2}))$ such that $[\tilde U_{\theta},C^A]=0$. In this case, the symmetry group becomes $U(1)\times \mathbb{Z}^T_2$, and the corresponding classification is $H^2(U(1)\times \mathbb{Z}^T_2, U_T(1))=\mathbb{Z}_2\times\mathbb{Z}_2$~\cite{Wen2012,Chen2013}. We stress that, in this example, anti-unitary symmetry is not necessary for the physical protection of zero modes. Even with only unitary symmetries, the topological classification remains non-trivial. Consequently, we focus primarily on unitary-symmetry protection, but will revisit anti-unitary symmetries in Sec.~\ref{sec:dybrking}, where dynamical symmetry breaking is discussed. The unitary-symmetry protection renders the zero modes robust against dynamical symmetry breaking and is essential for well-defined braiding results.


\section{Braiding statistics of zero modes}\label{sec:brazm}
In this section, we study the braiding statistics of topological zero modes in the one-dimensional bosonic SPT phase introduced above. Based on the symmetry analysis, we derive the effective Hamiltonian of the braiding process. A central prediction is that the reality condition of Hamiltonian reveals two distinct classes of braiding statistics. We propose two types of tri-junction-based braiding schemes to realize both classes of statistics. The first scheme involves a uniformly half-filled tri-junction, while the second introduces a controllable local defect with empty filling at the tri-junction crossover. Our findings apply to a wide range of bosonic SPT phases with real Hamiltonian realizations

\subsection{General theory of the braiding statistics}\label{sec:effHam}

Braiding statistics cannot be implemented in a single one-dimensional chain. To address this, we employ a tri-junction configuration, which allows for the controlled braiding of zero modes. The tri-junction consists of two or three spin chains connected at a point, effectively providing an extra spatial dimension that prevents quasiparticle collisions during the braiding process. Details of the tri-junction configuration are presented in the next subsection. The braiding is an unitary evolution which can be characterized by the evolution operator
\begin{equation}\label{eq:prop}
U(T) = \hat{T}e^{-i\int_0^T H(t)dt},
\end{equation}
where $\hat T$ denotes the time-ordered integral for the Hamiltonian $H(t)$ of the tri-junction, and $T$ is the total braiding time. The braiding evolution can further be characterized through an effective Hamiltonian defined as
\begin{eqnarray}
H_{E}&\equiv&\frac{i}{T}\text{log}U(T).
\end{eqnarray}
One can easily verify that the unitary symmetries of $H(t)$ at any $t$ are also preserved for the effective Hamiltonian $H_E$. These symmetries constrain the form of $H_E$. Furthermore, the braiding operators for different zero modes must satisfy the Yang–Baxter equation
\be
U_{12}U_{23}U_{12} = U_{23}U_{12}U_{23}.
\ee
where $U_{i(i+1)}$ denotes the counterclockwise braiding transformation between the $i$th and $(i+1)$th zero modes. For the spin-exchange model, we write down the most generic form of the effective Hamiltonian
\begin{eqnarray}\label{eq:efth}
H_{E}=\lambda\bigr(e^{i\alpha}\beta_{1}^{\dagger}\beta_{2} + h.c.\bigr)+\delta \left(n_1 + n_2 - 2n_1 n_2\right),
\end{eqnarray}
where the real parameters $\lambda$, $\delta$, and $\alpha$ will be determined by the symmetry constraint and generic fundamental features of the braiding process, and $\beta^{\dagger}_1$ ($\beta^{\dagger}_2$) denotes the left (right) zero-mode creation operator, with the particle number operator defined by $n_1 = \beta_1^\dagger \beta_1$ ($n_2 = \beta_2^\dagger \beta_2$). Substituting $H_E$ into the Yang-Baxter equation, we obtain the general form of braiding operator after some algebra (see Methods.~\ref{sec:yangbaxter} for details):
\begin{eqnarray} \label{eq:gebrop}
   U=I&+&ie^{-i\delta T}\left(e^{i\alpha}\beta_1^{\dagger}\beta_2+e^{-i\alpha}\beta_2^{\dagger}\beta_1\right)\nonumber\\
   &-&n_1-n_2+2n_1n_2.
\end{eqnarray}
For the real Hamiltonian $H(t)$, the zero modes after braiding must keep to be real. This enforces that under a full braiding, the transformation may generally take two possibilities that $U^2=\pm I$, which corresponds to the two possible Berry phases $0$ or $\pi$. Under this constraint the parameters $\delta$ and $\alpha$ are quantified into two sets of discrete values, giving the two distinct classes of braiding statistics which we show below in detail. The two classes of braiding statistics emerge as a universal feature of the bosonic SPT phases, and constitute a fundamental dynamical observable for a broad class of bosonic SPT phases with real Hamiltonians.

We return to the present specific spin-exchange model. For the first class with $U^2=I$, one has $\delta =\frac{2n\pi}{T}+\frac{\pi}{2T}$ with $n\in \mathbb{Z}$ and $\alpha=0$, and the second-quantized form of the braiding operator is (see Methods.~\ref{sec:yangbaxter})
\begin{equation}\label{eq:convtrans}
    U^{\text{conv}}=I+\beta_{1}^{\dagger}\beta_{2}+\beta_{2}^{\dagger}\beta_{1}-n_1-n_2+2n_1 n_2.
\end{equation}
Here we use the superscript ``conv" to denote ``conventional" that will be clarified below. The above braiding transforms the zero
modes as
\begin{equation}\label{eq:tribr}
U^{\text{conv}}\beta_{1}(U^{\text{conv}})^{\dagger} =  \beta_{2}, \
U^{\text{conv}}\beta_{2}(U^{\text{conv}})^{\dagger} =  \beta_{1},
\end{equation}
which follows the standard bosonic commutation relation of the two bosonic zero modes. This is in contrast to the braiding of MZMs, in which a minus sign appears since the Majorana zero-mode operators obey fermion commutation relation.

The more intriguing prediction is that the bosonic zero modes allow a second class of exotic unconventional braiding transformations, corresponding to $\delta =\frac{2n\pi}{T}$ with $n\in \mathbb{Z}$ and $\alpha=\frac{\pi}{2}$ in Eq.~\eqref{eq:gebrop}, given by the braiding operator (see Methods.~\ref{sec:yangbaxter}):
\begin{eqnarray}\label{eq:deftrans}
U^{\text{def}}&=&I-\beta_{1}^{\dagger}\beta_{2}+\beta_{2}^{\dagger}\beta_{1}-n_1-n_2+2n_1 n_2 \nonumber
\\
&=&\exp\left[\frac{\pi}{2}(\beta_{2}^{\dagger}\beta_{1} - \beta_{1}^{\dagger}\beta_{2})\right].
\end{eqnarray}
Here the superscript ``def" denotes ``defect-assisted", as we shall see in the next subsection that the unconventional braiding statistics can be achieved with assistance by a controlled defect. Under this braiding operation, the zero modes transform as
\begin{equation}\label{eq:optransnonlinear}
\begin{aligned}
U^{\text{def}}\beta_{1}(U^{\text{def}})^{\dagger} =\beta_{2}-2\beta_{1}^{\dagger}\beta_{1}\beta_{2}, \\
U^{\text{def}}\beta_{2}(U^{\text{def}})^{\dagger} =  -\beta_{1}+2\beta_{2}^{\dagger}\beta_{2}\beta_{1},
\end{aligned}
\end{equation}
and the full braiding recovers a $\pi$ phase factor for each zero mode $(U^{^{\text{def}}})^2 \beta_{1(2)} ((U^{\text{def}})^\dagger)^2 = -\beta_{1(2)}$, rendering $U^2=-I$. This nonlinear braiding operation is a fractionalization of the conventional braiding statistics in Eq.~\eqref{eq:optransnonlinear}.
The nonlinear transformations is also a hallmark feature of fractionalization in correlated systems, similar to those hosting parafermion zero modes~\cite{Hong2024}.
To further show the nontrivial essence of the unconventional braiding, we map $U^{\text{def}}$ to the braiding of two pairs of symmetry-protected MZMs~\cite{liu2014non,gao2016symmetry,hong2022unitary,Haim2019Time,Masaki2024}. This structure can be elucidated via a Jordan-Wigner transformation on the zero modes 
\begin{eqnarray}\label{eq:optransnonlinear2}
\beta_{1}=\frac{1}{2}(\gamma_1+i\tilde\gamma_1), \ \beta_{2}=\frac{i}{2}e^{\frac{\pi}{2}\tilde\gamma_1\gamma_1}(\gamma_{2}+i\tilde\gamma_{2}),
\end{eqnarray}
which maps the hard-core bosonic zero modes to Majorana modes $\gamma_{1}$, $\gamma_2$, $\tilde\gamma_1$ and $\tilde\gamma_{2}$. We further obtain
\begin{eqnarray}\label{eq:MajoranaPair}
U^{\text{def}}=\exp\bigr(-\frac{\pi}{4}\tilde\gamma_1\tilde\gamma_2\bigr)\exp\bigr(-\frac{\pi}{4}\gamma_1\gamma_2\bigr).
\end{eqnarray}
This novel equivalence demonstrates that the unconventional braiding manifests the emergent fermionic zero-mode behavior. While the Majorana modes transform linearly under braiding, the transformation in Eq.~\eqref{eq:optransnonlinear} related to the Majorana braiding through Jordan-Wigner mapping naturally contains nonlinear terms.

The braiding transformations on the ground states can be obtained straightforwardly. Owing to particle-number conservation, the ground-state space separates into three sectors characterized by the total zero-mode occupation number as $n_1+n_2=0,1,2$. The braiding operators act nontrivially on the half-filled subspace sector ($n_1+n_2=1$) that $U^{\text{def}}|1,0\rangle=|0,1\rangle$ and $U^{\text{def}}|0,1\rangle=-|1,0\rangle$, while for the conventional braiding we have $U^{\text{conv}}|1,0\rangle=|0,1\rangle$ and $U^{\text{conv}}|0,1\rangle=|1,0\rangle$. In comparison, the two states $|00\rangle$ and $|11\rangle$ are invariant under both braiding transformations. For $U^{\text{def}}$ this invariance is nontrivial and ensured by the nonlinear terms in Eq.~\eqref{eq:optransnonlinear}.
We point out that the unconventional braiding transformation $U^{\text{def}}$ on the four degenerate states is the same as the braiding transformation of two Majorana pairs on the four ground states in the symmetry-protected topological superconductors~\cite{liu2014non,hong2022unitary}. This also explains why the unconventional braiding can be mapped to the SPNA statistics of Majorana pairs as shown above. 


The existence of two classes of SPNA statistics 
facilitates to realize non-commuting topological gates in the single-qubit Hilbert space. The $2$D half-filled subspace $\mathcal{H}_0=\text{span}\{\ket{1,0},\ket{0,1}\}$ naturally serves as the qubit space, on which two remarks are given. First, as explained below, the subspace $\mathcal{H}_0$ encodes a nonlocal qubit under the symmetry protection. Second, the two braiding operators in Eqs.~\eqref{eq:convtrans} and \eqref{eq:deftrans} in this subspace yield $\sigma^x$ and $-i\sigma^y$ gates, respectively, rendering the non-commutable operations in the single quibt setting. 
For simplicity, we relabel the two basis states as
\begin{equation}
   \ket{\beta_1}\equiv\ket{1,0} \, \text{and} \, \ket{\beta_2}\equiv\ket{0,1}.
\end{equation}
In comparison, as is well-known for the MZMs, a single qubit can be encoded by four zero modes, and the non-commuting gates are realized by braiding different pairs of the four Majorana modes~\cite{ivanov2001non,sarma2005topologically,alicea2011non}. Here in our setting two hard-core bosonic zero modes are sufficient to encode one qubit and realize non-commutable gates.

We conclude this subsection with a few important remarks on the generic theory for the SPNA statistics of bosonic zero modes in the bosonic SPT phases. In general, the non-Abelian statistics of quasiparticles are obtained when the following three conditions are satisfied~\cite{nayak2008non}. First, the presence of quasiparticles leads to ground state degeneracy, namely, the quantum dimension of the quasiparticles is greater than one. Second, the unitary transformation between two different degenerate ground states cannot be achieved by local operations, but can be achieved by nonlocal braiding of spatially separated quasiparticles. Third, the braiding operations are topological (discrete). It is clear that the conditions I and III are satisfied by the bosonic zero modes in the present SPT phases. The second condition is regarding the nonlocality of the quasiparticle states. We note that, in general, a non-local state is defined as the one that cannot be changed by local operators {\em under certain conditions}, but can be transformed by nonlocal operations. In the framework of Majorana modes, the non-locality of the states refers to that such states cannot be changed by the local operations respecting the topological gap. Actually, if the local perturbation is dynamical, for instance, for the local dynamical noise which couples the local Majorana mode to the bulk, the effective nonlocal couplings can be resulted even only local perturbations are introduced, and the topological quibts will not be stable in general~\cite{MajoranaNoise}. For the qubits defined on zero modes in the present bosonic SPT phase, the key constrain is the symmetry protection. Even though each single zero mode is a local state, as long as the unitary symmetry is preserved, the two-mode qubit states, which define the ground state degeneracy, cannot be changed by any local operators. In other words, it is the symmetry protection that enforces that the ground states become nonlocal and cannot be changed by any local transformation, but can be transformed by braiding. As a result, all the three conditions are satisfied by the bosonic zero modes of the present SPT phases, rendering the rigorous theory of the SPNA statistics uncovered here. 

The two classes of non-Abelian braiding statistics of the bosonic zero modes also update the basic understanding of the bosonic SPT phases. For the fundamental theory of bosonic SPT phases, research has mainly focused on developing the topological classification and characterization of these phases through the topological invariants, bulk-boundary correspondence, and symmetry protection mechanisms, etc. While the bosonic topological boundary modes have been widely studied, the braiding statistics of such modes were not considered before. The present prediction shows that, due to symmetry protection, the braiding statistics of topological zero modes in bosonic SPT phases are highly nontrivial and generally fall into two distinct classes, one of which is a fractionalization of the other in real-Hamiltonian systems. The presence of the two distinct classes of SPNA statistics is also a fundamental nature of the bosonic zero modes in the correlated systems, providing a fundamental dynamical character of the bosonic SPT phases.

\begin{figure}[t]
    \centering
    \includegraphics[scale=0.25]{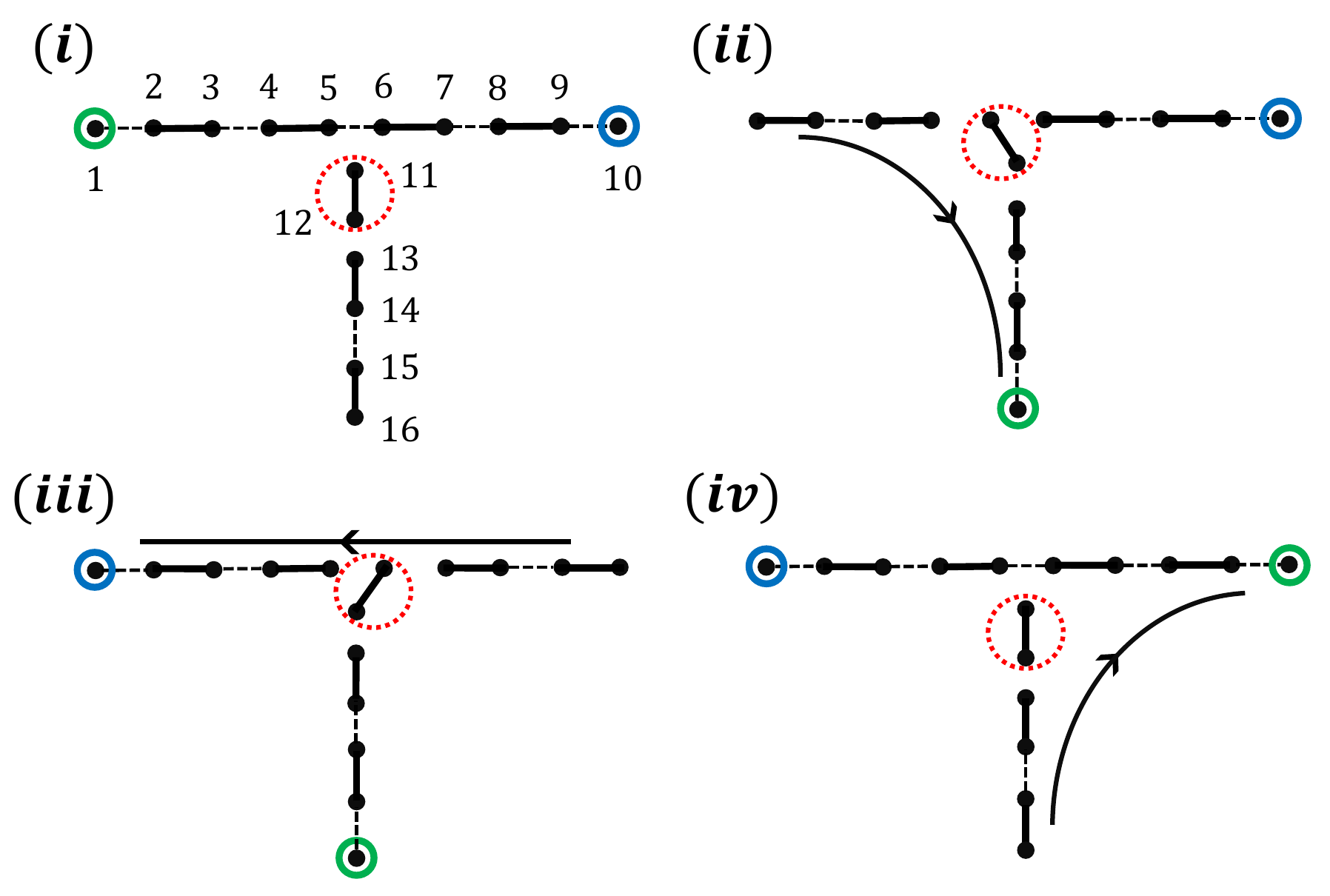}
    \caption{Configuration changes of the local defect (enclosed by a red dashed line) during the defect-assisted braiding process.}
    \label{fig:juncconfig}
\end{figure}

\subsection{Braiding schemes with tri-junctions}\label{sec:bztj}

While the generic theory predicts two distinct classes of non-Abelian statistics, devising explicit schemes to realize them remains highly nontrivial. In this subsection we propose two types of braiding schemes to realize the two distinct classes of braiding statistics. In particular, the conventional braiding statistics can be realized with a standard tri-junction
depicted in Fig.~\ref{fig:specandbraiding}(c), which consists of three bosonic SPT chains intersecting at a central site. The entire system is uniformly half-filled. To realize unconventional braiding statistics, we propose a defect-assisted tri-junction illustrated in Fig.~\ref{fig:specandbraiding}(d). The defect is located at the junction point and formed by two sites, connecting the horizontal and vertical topological chains. In contrast to the half-filling condition in the topological chains, the local defect is set to be either empty or fully filled (both yielding identical results). We find that the defect leads to a nontrivial non-Abelian Berry phase for the zero modes transporting through the tri-junction of this correlated system, for which the unconventional braiding statistics is realized. 

Both tri-junctions, with initial coupling configurations shown in Figs.~\ref{fig:specandbraiding}(c) and \ref{fig:specandbraiding}(d), host two zero modes localized at the horizontal ends of the junction, which induce the ground-state degeneracy. The braiding of zero modes is achieved by adiabatically moving their localization positions, as illustrated in Fig.~\ref{fig:specandbraiding}(c) [or Fig.~\ref{fig:specandbraiding}(d)]. The braiding process begins by moving the left zero mode to the bottom of the tri-junction, followed by transporting the right zero mode to the left end, and finally the bottom zero mode to the right end. This procedure is decomposed into a sequence of elementary steps, each shifting the zero mode by two sites, as depicted in Fig.~\ref{fig:specandbraiding}(e). The dynamics of the elementary moving step involving bosonic modes $b_{1,2,3}$ are governed by the Hamiltonian
\begin{equation}\label{eq:emp}
H_{\text{e}}(t)=u(t)b_{1}^{\dagger}b_{2}+v(t)b_{2}^{\dagger}b_{3}+h.c.,
\end{equation}
where $u(t)$ and $v(t)$ are time-dependent real parameters. Initially, the parameters satisfy $u(0)<v(0)$, localizing the zero mode at site $1$. By adiabatically tuning the parameters to $u(T_{0})>v(T_{0})$, with $T_0$ as the duration of the elementary step, we move the zero mode from site $1$ to site $3$. Adiabaticity requires the time $T_0$ to be sufficiently large compared to the inverse of the energy gap. The time dependence of $u(t)$ and $v(t)$ can take various forms such as linear functions or trigonometric functions. In this work, we adopt a smooth exponential function~\cite{Boross2019}. The parameters $u(t)$ and $v(t)$ change over time as
\begin{equation}\label{eq:evofunc}
\begin{aligned}
u(t)=&v_{\text{min}} + (v_{\text{max}}-v_{\text{min}})\chi(t/T_0),
\\
v(t)=&v_{\text{max}}-(v_{\text{max}}-v_{\text{min}})\chi(t/T_0),
\end{aligned}
\end{equation}
where
\begin{equation}\label{eq:timeevofunc}
\chi(t)=\frac{e^{-\frac{1}{t}}}{e^{-\frac{1}{1-t}}+e^{-\frac{1}{t}}},
\end{equation} and $v_{\text{min}}$ ($v_{\text{max}}$) is the minimum (maximum) value of the coupling strengths during the process.

For the first type of braiding scheme, the tri-junction in Fig.~\ref{fig:specandbraiding}(c) is in the ground state of a half-filled bosonic SPT system. Due to particle-hole symmetry, the dynamics of ground states $\ket{\beta_1}$ and $\ket{\beta_2}$ are identical, resulting in no net geometric phase difference between $\ket{\beta_1(T)}$ and $\ket{\beta_2(T)}$. The ground states $\ket{\beta_1}$ and $\ket{\beta_2}$ transform as follows after braiding:
\begin{equation}\label{eq:statetranstr}
 \ket{\beta_1} \to \ket{\beta_2}, \quad \ket{\beta_2} \to  \ket{\beta_1},
\end{equation}
where the overall dynamical phase, identical for $\ket{\beta_1(T)}$ and $\ket{\beta_2(T)}$, is neglected. This scheme realizes conventional braiding statistics of hard-core bosons, with operator transformations given by Eq.~\eqref{eq:tribr}. We refer to this scheme as {\it conventional braiding}.

We now consider the scheme for the second class of braiding statistics, as illustrated in Fig.~\ref{fig:specandbraiding}(d), where the half-filled horizontal chain supports two zero modes at the ends, together with an empty local defect at the junction. During the braiding, the local defect changes its position around the junction, while remaining isolated from the remaining part of the tri-junction at the termini, and returns to the original state after a single braiding [Fig.~\ref{fig:juncconfig}]. This fulfills the criteria for the well-defined braiding operations. Braiding the two zero modes in the presence of a local defect induces a $\pi$ phase difference between $\ket{\beta_1(T)}$ and $\ket{\beta_2(T)}$. This phase is topological in origin, as the defect is located at the geometric singularity of the tri-junction and alters the topology of the braiding process. As shown in Methods~\ref{sec:berry}, we compute the non-Abelian Berry phases in a tri-junction setting. In particular, we show that the two classes of statistics fall into distinct topological classes, the classification of which can be related to the Stiefel--Whitney classes of real vector bundles. The defect at the junction acts as a singular point, and leads to a topological Berry phase $e^{i\Delta\phi}=\langle\beta_2|U(T)|\beta_1\rangle\langle\beta_1|U(T)|\beta_2\rangle$, with $\Delta\phi=\pi$ for the unconventional braiding statistics. Supplementary~\ref{sec:brdimer} presents an analytic proof of the braiding results for larger size tri-junctions in the dimerized limit, while the topological $\pi$ phase persists away from the dimerized limit. With the topological $\pi$ phase, the ground states $\ket{\beta_1}$ and $\ket{\beta_2}$ transform as
\begin{equation}\label{eq:statetransntr}
 \ket{\beta_1} \to \ket{\beta_2}, \quad \ket{\beta_2} \to  -\ket{\beta_1},
\end{equation}
up to an overall dynamical phase. The minus sign in the above transformation implies an emergent fermionic type of statistics, which aligns with the operator transformation in Eq.~\eqref{eq:optransnonlinear}. We refer to this novel braiding scheme as {\it defect-assisted braiding}.

It is convenient to represent the braiding results as matrices in the following basis:
\begin{equation}\label{eq:qubitbasis}
\ket{\xi} \equiv \frac{1}{\sqrt{2}}(\ket{\beta_1}+\ket{\beta_2}), \quad \ket{\eta} \equiv \frac{1}{\sqrt{2}}(\ket{\beta_1}-\ket{\beta_2}).
\end{equation}
According to the analysis above, in the adiabatic limit ($T_0 \rightarrow +\infty$), the braiding of zero modes gives rise to two distinct matrices (up to a global dynamical phase):
\begin{eqnarray}\label{eq:braidingab}
&\text{lim}_{T_0 \rightarrow +\infty}
\left(
\begin{array}{cc}
    \langle \xi | \xi(T) \rangle & \langle \xi | \eta(T) \rangle \\
    \langle \eta | \xi(T) \rangle & \langle \eta | \eta(T) \rangle
\end{array}
\right)\bigg|_{T} \nonumber
\\
&= \left(
\begin{array}{cc}
    0 & 1 \\
    -1 & 0
\end{array}
\right) \quad \text{or} \quad \left(
\begin{array}{cc}
    1 & 0 \\
    0 & -1
\end{array}
\right),
\end{eqnarray}
in the case with or without the local defect. Here, $\ket{\xi(T)} \equiv U(T) \ket{\xi}$ and $\ket{\eta(T)} \equiv U(T) \ket{\eta}$.
In this new single-qubit basis $\{\ket{\xi}, \ket{\eta}\}$, we achieve the non--commutative braiding matrices $\sigma^z$ for conventional braiding and $i\sigma^y$ for defect-assisted braiding.

\begin{figure*}[t]
    \centering
    \includegraphics[scale=0.25]{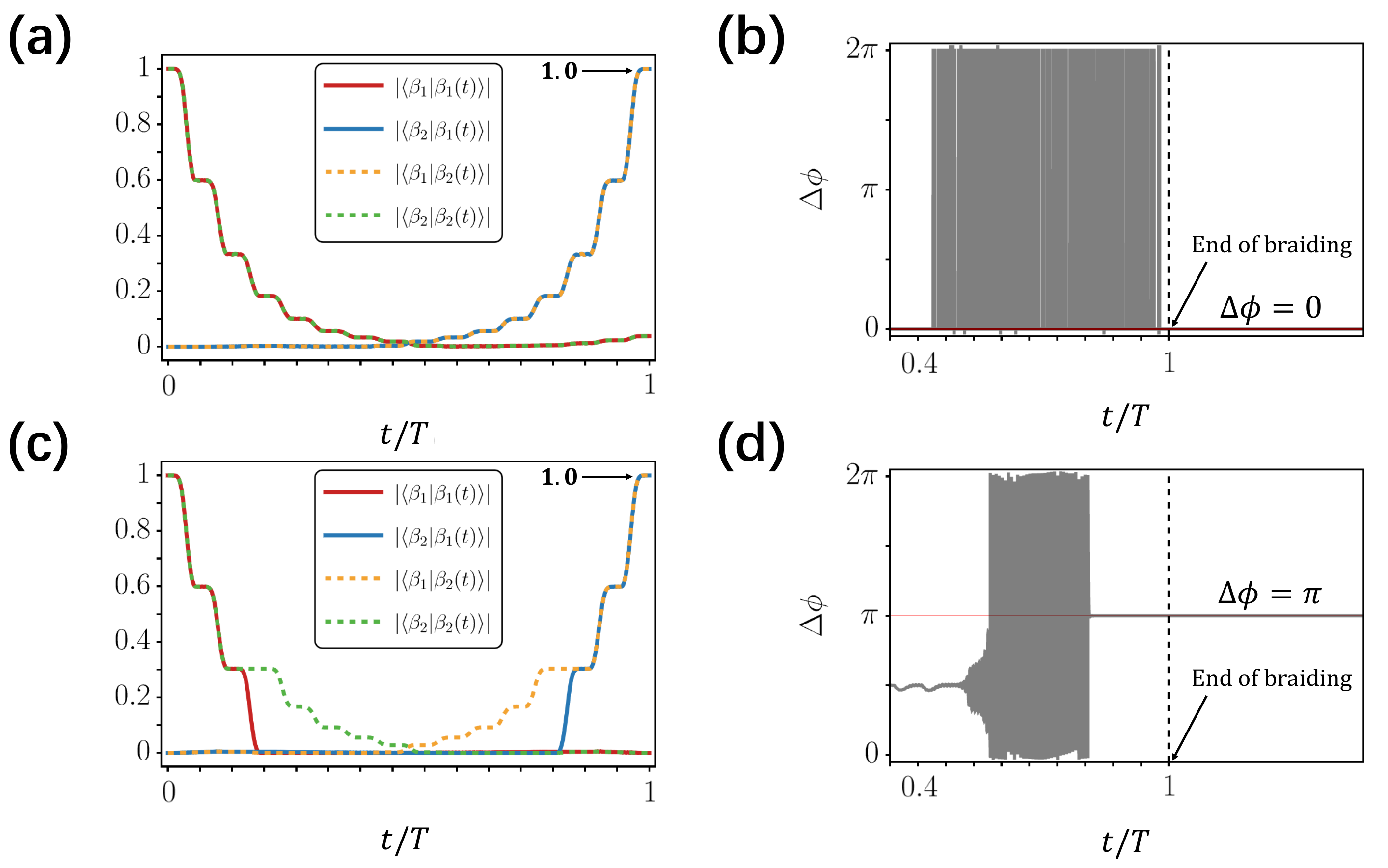}
    \caption{Numerical time evolution of zero-mode wavefunctions for two braiding schemes. (a) Amplitudes of and (b) phase difference between $\ket{\beta_1(t)}$ and $\ket{\beta_2(t)}$ for the first type of braiding scheme (conventional braiding). (c) Amplitudes and (d) phase difference for the second type of braiding scheme (defect-assisted braiding). The horizontal axis represents the evolution time in units of the total braiding time $T$. The strong and weak couplings take $v_{\text{max}}=1$ and $v_{\text{min}}=0.1$. The simulations use a precision of time step $\Delta t = 0.1$ and an elementary step time $T_0 = 60$ for the first scheme ($T_0 = 36$ for the second). Each braiding consists of 15 elementary steps of zero-mode motion, giving a total braiding time $T = 15T_0 = 900$ for the first scheme ($T = 15T_0 = 540$ for the second). The red horizontal line denotes the theoretical prediction for $\Delta \phi = \mathrm{Arg}(\langle \beta_2 | \beta_1(t) \rangle) - \mathrm{Arg}(\langle \beta_1 | \beta_2(t) \rangle) \, \mathrm{mod} \, 2\pi$ after braiding. The vertical dashed line marks the end of the braiding, after which the system is kept static to illustrate the stabilized value of $\Delta \phi$, as shown in the figure. }
    \label{fig:trijuncwave}
\end{figure*}

To leverage the non-Abelian nature of braiding transformations, one can extend the system to multiple copies for encoding logical qubits and implementing single- and two-qubit gates via braiding operations. For example, consider a system with two copies, hosting four zero modes $\beta_{1 \cdots 4}$. The subspace where only a single zero mode is occupied forms a $4$-dimensional space, allowing the following encoding for logical qubits:
\begin{equation}
|\overline{00}\rangle = |\emptyset\rangle |\xi\rangle, \, |\overline{10}\rangle = |\xi\rangle |\emptyset\rangle, \, |\overline{01}\rangle = |\emptyset\rangle |\eta\rangle, \, |\overline{11}\rangle = |\eta\rangle |\emptyset\rangle.
\end{equation}
Here $\ket \emptyset$ denotes the state with no zero mode occupied in the tri-junction. Under this specific encoding, the CZ gate and the CNOT gate are realized by
\begin{equation}\label{eq:CNOT}
    \text{CZ gate: } U_{12}^{\text{conv}}, \quad \text{CNOT gate: } U_{12}^{\text{conv}} U_{12}^{\text{def}},
\end{equation}
The single-qubit Pauli gates, embedded in the two-logical qubit space, can be implemented as follows:
\begin{eqnarray}\label{eq:singlequbit}
\overline{I \otimes Z} &=& U^{\text{conv}}_{12} \otimes U^{\text{conv}}_{34}, \nonumber
\\
\overline{I \otimes X}  &=& (U^{\text{conv}}_{12} \otimes U^{\text{conv}}_{34})(U^{\text{def}}_{12}\otimes U^{\text{def}}_{34}),
\\
\overline{X \otimes I} &=& U^{\text{conv}}_{23}U^{\text{conv}}_{12}U^{\text{conv}}_{34}U^{\text{conv}}_{23}, \nonumber
\\
\overline{Z \otimes I} &=& (U^{\text{def}}_{12})^{2}. \nonumber
\end{eqnarray}
We see that the two classes of statistics, contrlled by the filling of local defect, facilitate the realization of quantum gates. Compared to gate realization with MZMs using dense encoding, our approach avoids the embedding problem associated with Majorana qubits~\cite{Ezawa2024,Ivanov2001,sarma2005topologically,Georgiev2006,Kraus2013,Lachezar2008,Ahlbrecht2009}.

\section{Numerical simulations}\label{sec:num}

We present in this section the numerical results to support the rigorous theory in the previous section. We demonstrate the time evolution of ground-state wavefunctions and compute the average fidelities of the braiding matrices for both conventional and defect-assisted braidings. In particular, we focus on the half-filled sector $\mathcal{H}_0$ spanned by $\ket{\beta_1}$ and $\ket{\beta_2}$, where the non-Abelian nature is prominently manifested.

We consider specific tri-junctions with $16$ sites depicted in Figs.~\ref{fig:specandbraiding}(c) and \ref{fig:specandbraiding}(d), which implement the first and second types of braiding scheme, respectively. For the first type of braiding scheme, the dynamics of the system are governed by the time-dependent Hamiltonian
\begin{eqnarray}\label{eq:hamfisrt}
H^{\text{tri}}_{\text{1st}}(t)&=&\sum_{i=1}^{4}v_{i,i+1}(t)b^{\dagger}_{i+1}b_{i}+\sum_{i=6}^{9}v_{i,i+1}(t)b^{\dagger}_{i+1}b_{i} \nonumber
\\
&&+\sum_{i=11}^{15}v_{i,i+1}(t)b^{\dagger}_{i+1}b_{i}+v_{5,11}b^{\dagger}_{11}b_{5} \nonumber
\\
&&+v_{6,11}b^{\dagger}_{11}b_{6}+h.c.,
\end{eqnarray}
where $v_{i,j}$ denotes the coupling strength between sites $i$ and $j$. For the second type of braiding scheme, the time-dependent junction Hamiltonian is
\begin{eqnarray}\label{eq:hamsecond}
H^{\text{tri}}_{\text{2nd}}(t)&=&\sum_{i=1}^{9}v_{i,i+1}(t)b^{\dagger}_{i+1}b_{i}+\sum_{i=11}^{15}v_{i,i+1}(t)b^{\dagger}_{i+1}b_{i} \nonumber
\\
&&+v_{5,11}b^{\dagger}_{11}b_{5}+v_{6,11}b^{\dagger}_{11}b_{6}+h.c.,
\end{eqnarray}
The Hamiltonians in Eqs.~\eqref{eq:hamfisrt} and \eqref{eq:hamsecond} are referred to as the first and the second type of tri-junction Hamiltonians, respectively. Initially, the coupling strengths are set according to the configurations in Figs.~\ref{fig:specandbraiding}(c) and \ref{fig:specandbraiding}(d), with the strong and weak couplings $v=1$ and $v'=0.1$. To investigate the effect of system size on gap protection for tri-junctions, we numerically calculate the energy gap $\delta E$ between the ground state and the first excited state of the first type of tri-junction Hamiltonian at $t=0$ as a function of tri-junction size, shown in Fig.~\ref{fig:specandbraiding}(f). The gap $\delta E$ is slightly larger for smaller sizes and rapidly approaches a constant value as the size increases. A similar behavior is observed for the second type of tri-junction Hamiltonian. For $t>0$, the energy gap $\delta E$ deviates slightly from a constant value and oscillates periodically with time. Nevertheless, the amplitude of the oscillation is very small, allowing the gap to be approximated as constant during the braiding process. Therefore, the proposed tri-junction configuration provides gap protection throughout the braiding process.

To implement the braiding, we employ the scheme in Fig.~\ref{fig:specandbraiding}(e) to manipulate zero modes. In each step, we take three sites, decrease the initial strong coupling strength $v$ and increase the weak coupling strength $v'$ over a time period $T_0$, following the functional dependence in Eq.~\eqref{eq:evofunc}, with $v_{\text{max}}=v$ and $v_{\text{min}}=v'$. This adiabatic tuning shifts the zero mode by two sites. The braiding process consists of $15$ moving steps. For the second type of braiding scheme, special care is needed when the zero mode passes near the local defect. The subtlety is to ensure that the defect remains isolated both before and after the zero mode passes through it. To achieve this, we first decouple the zero mode from the chain as it moves adjacent to the defect, ensuring that it is localized on a single site. We then exchange the zero mode with the local defect in a dimerized-limit fashion. Finally, we recouple the zero mode to the chain and complete the remaining movements. This approach guarantees that the local defect returns to its original state after braiding. The critical aspect of our method is maintaining defect isolation by decoupling the zero mode when it passes through. In Supplementary~\ref{sec:effError}, we quantitatively analyze errors that arise when defect is not fully isolated, confirming that our scheme effectively prevents such errors.

\begin{figure*}[t]
    \centering
    \includegraphics[scale=0.25]{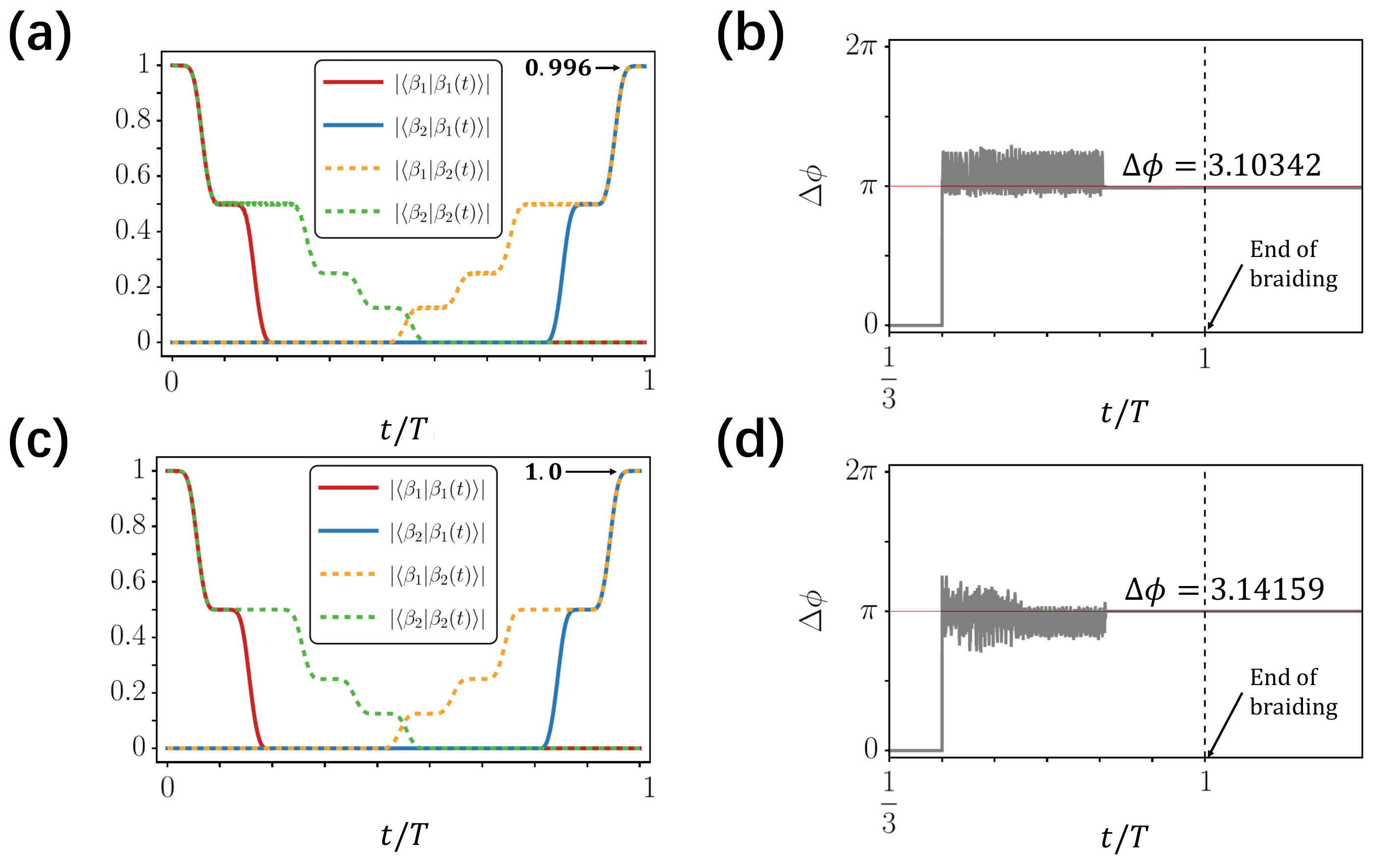}

    \caption{Numerical time evolution of zero-mode wavefunctions under random hopping disorder in the second braiding scheme (defect-assisted braiding) using a dimerized tri-junction. (a) and (b) show the evolution of the amplitudes of $\ket{\beta_1(t)}$, $\ket{\beta_2(t)}$, and their phase difference under complex-valued random hopping disorders. (c) and (d) show the evolution under real-valued random hopping disorders. The calculations are performed on the dimerized tri-junctions with $10$ sites [Fig.~\ref{fig:zeromode}(b)].The horizontal axis shows the evolution time in units of the total braiding time $T$. The strong and weak couplings take $v_{\text{max}}=1$ and $v_{\text{min}}=0$ Simulations are performed with a time step $\Delta t = 0.1$ and an elementary step time $T_0 = 80$. Each braiding consists of $9$ elementary steps of zero-mode motion, giving a total time $T = 9T_0 = 720$. The red horizontal line represents the theoretical prediction of $\Delta \phi = \text{Arg}(\langle \beta_2 | \beta_1(t) \rangle) - \text{Arg}(\langle \beta_1 | \beta_2(t) \rangle) \, \text{mod} \, 2\pi$ in the absence of dynamical symmetry breaking. The vertical dashed line marks the end of the braiding process, after which the system is held static to demonstrate the stabilized value of $\Delta \phi$. The values of $\Delta \phi$ for both disorder cases are indicated in the figure, and the numerical results are averaged over $100$ random disorder realizations. The disorder strength is introduced as a fluctuation of the exchange coupling, $\delta v_i = s_i v_i$, where $s_i$ follows a Gaussian distribution with zero mean and standard deviation $\sigma = 0.05$.}
    \label{fig:disorderwave}
\end{figure*}

The total braiding time is $T=15T_0$. We compute the evolution operator $U(T)$ by discretizing the braiding process into small time intervals $\Delta t$ and iteratively applying the time-evolution operator at each step:
\be
U(T) \approx \text{lim}_{\Delta t\rightarrow 0}e^{-iH(T)\Delta t} e^{-iH(T-\Delta t)\Delta t} \cdots e^{-iH(0)\Delta t}.
\ee
The time $T_0$ is chosen to be large enough to ensure adiabaticity.  Fig.~\ref{fig:trijuncwave} shows the time evolution of ground-state wavefunctions for both braiding schemes. From $t=0$ to $t=T$, the states $\ket{\beta_1}$ and $\ket{\beta_2}$ interchange, as illustrated by the amplitude results in Figs.~\ref{fig:trijuncwave}(a) and \ref{fig:trijuncwave}(c). To obtain the braiding matrix, we calculate the phase difference $\Delta \phi(t)$ defined by
\begin{equation}\label{eq:deltaphi}
    \Delta \phi(t) = \text{Arg}(\langle \beta_2 | \beta_1(t) \rangle )-\text{Arg}(\langle \beta_1 | \beta_2(t)\rangle) \, \text{mod} \, 2\pi.
\end{equation}
The results in Figs.~\ref{fig:trijuncwave}(b) and \ref{fig:trijuncwave}(d) show that the phase differences are $0$ (for the first type of braiding scheme) and $\pi$ (for the second type of braiding scheme), corresponding to the transformation $\ket{\beta_1} \rightarrow \ket{\beta_2}$ and $\ket{\beta_2} \rightarrow \ket{\beta_1}$ for conventional braiding, and $\ket{\beta_2} \rightarrow -\ket{\beta_1}$ for defect-assisted braiding. The results confirm that the braiding matrices are $\sigma^z$ for the conventional braiding and $i\sigma^y$ for the defect-assisted braiding in the $\{\ket{\xi},\ket{\eta}\}$ basis, up to an overall dynamical phase.

To further quantify the accuracy of the numerical results, we calculate the average fidelity $F$ of the braiding matrix (see Methods~\ref{sec:fidelity} for the definition). For defect-assisted braiding, the average fidelity is $F(U, i\sigma^y) = 0.99999999$, while for conventional braiding, it is $F(U, \sigma^z) = 0.999$. The decrease in fidelity for conventional braiding is due to the small energy splitting $\Delta E$ between two ground states caused by finite-size effects [Fig.~\ref{fig:specandbraiding}(b)]. In contrast, defect-assisted braiding is less affected by the finite-size effect. These results show that there is no need to tune the system to the dimerized limit to achieve high fidelity. The scheme exhibits substantial robustness against finite-size effects.

\begin{figure*}[t]
    \centering
    \includegraphics[scale=0.24]{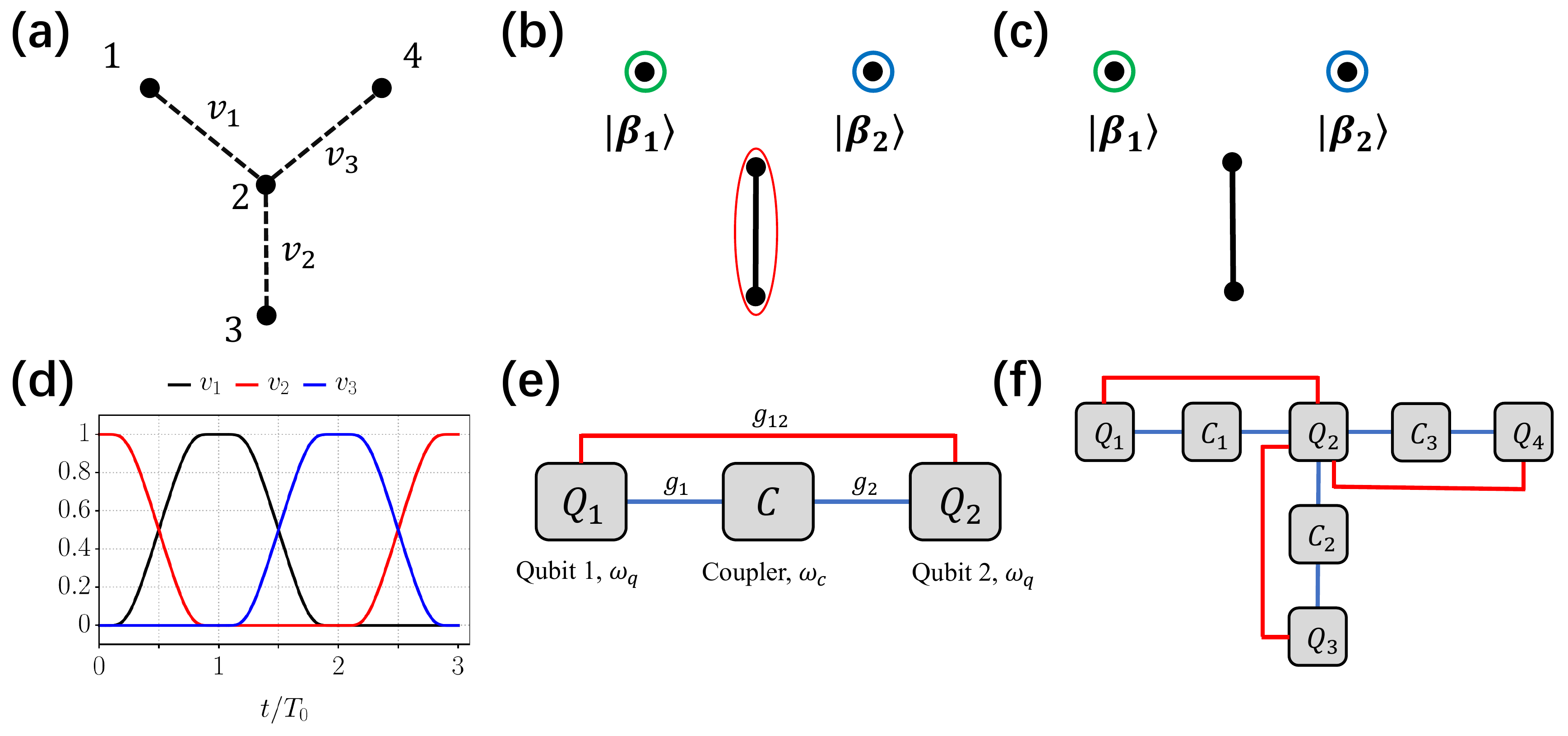}
    \caption{Illustration of the experimental scheme. (a) The minimal model with four spins. Spins $2$ and $3$ collectively serve as a defect. Spins $1$ and $4$ support zero modes in the dimerized limit. The dashed lines represent spin-exchange couplings. (b) and (c) Initial configurations for the first (conventional braiding) and second (defect-assisted braiding) type of braiding schemes. A red oval encircling two spins represents the bell state $\frac{1}{\sqrt{2}}(\ket{10}-\ket{01})$. (d) The time-dependent coupling strengths as functions of time to achieve braiding process. The coupling strength is in unit of its maximum value $v$. The ascending (descending) slope takes the form $\chi(t/T_0)$ ($1-\chi(t/T_0)$). (e) Tunable qubit-qubit interaction via a coupler, where $\omega_q$ and $\omega_c$ are the qubit and coupler frequencies, respectively, and $g_1$, $g_2$ and $g_{12}$ are fixed nearest neighbour and next nearest neighbour couplings. (f) Superconducting qubit tri-junction with tunable qubit-qubit interactions. Red and blue lines indicate the direct qubit-qubit coupling and qubit-coupler coupling, respectively.}
    \label{fig:expconfig}
\end{figure*}

Finally we comment the analogy nature of the simulation of the quasiparticles with finite-size topological systems. 
As shown in Fig.~\ref{fig:specandbraiding}(f) and mentioned above, the gap decreases slightly versus system size and approaches a constant quickly as the size increases. During the braiding process, the gap oscillates weakly around the constant. This tells that the gap protection in a finite system is nearly identical to that in the thermodynamic limit. In consequence, the SPNA statistics for the topological phases defined in the thermodynamic limit can be realized in the finite-size system with the same gap protection. In other words, the increase of the system size shall also not significantly increase the complexity of the simulation of the SPNA statistics, marking a key difference from the digital simulation. Finally, thanks to the gap protection, the experimental realization of the SPNA statistics can be achieved with the minimal tri-junction configuration, a strategy that we adopt in Sec.~\ref{sec:experm}.

\section{Symmetry protection mechanism of the braiding statistics}\label{sec:dybrking}

In this section, we further investigate the differences between unitary and anti-unitary symmetries in the symmetry protection mechanism of SPNA statistics in bosonic SPT phases. We show that SPNA statistics is robust against errors that preserve the unitary symmetries, as exemplified by real random disorder in the coupling strengths. By contrast, complex random disorder in the coupling strengths, although preserving the anti-unitary symmetry, destroys the SPNA statistics through dynamical symmetry breaking.

For convenience, we consider the second type of braiding scheme (defect-assisted braiding) in the dimerized limit (dynamical symmetry breaking for the conventional braiding is similar). We introduce random imaginary hopping terms:
\begin{equation}
H_p(t) =  i\sum_{i=1,2,...,6,8,9}\delta v_{i}(t) b^{\dagger}_{i+1}b_{i}+i\delta v_7(t)b^{\dagger}_{8}b_{4}+h.c.,
\end{equation}
to the Hamiltonian of the junction (Eq.~\eqref{eq:hamevo}), where $\delta v_i(t) = s_{i} v_i(t)$ and each $s_i$ is a Gaussian random variable with zero mean and standard deviation $\sigma=0.05$. This term preserves anti-unitary symmetry generated by $C^A$ and the particle number conservation, but breaks the unitary particle-hole symmetry. In Figs.~\ref{fig:disorderwave}(a) and \ref{fig:disorderwave}(b), we plot the evolution of amplitudes of zero-mode states $\ket{\beta_1}$, $\ket{\beta_2}$ and their phase difference under the Hamiltonian $H(t)+H_p(t)$. Fig.~\ref{fig:disorderwave}(a) shows that  $|\langle\beta_1|\beta_2(t=T)\rangle|$ and $|\langle\beta_2|\beta_1(t=T)\rangle|$ reach $0.996$, while $|\langle\beta_1|\beta_1(t=T)\rangle|$ and $|\langle\beta_2|\beta_2(t=T)\rangle|$ are $0$, indicating that the states run out of the subspace $\mathcal{H}_0$. Fig.~\ref{fig:disorderwave}(b) shows that the phase difference $\Delta \phi$ stabilizes at the value deviating from $\pi$ after braiding, confirming that the non-unitary symmetry undergoes dynamical breaking, causing the braiding statistics to become ill-defined in the ground-state subspace. In comparison, Figs.~\ref{fig:disorderwave}(c) and \ref{fig:disorderwave}(d) show the evolution under the Hamiltonian $H(t) - i H_p(t)$, which represents real random hopping disorders and preserves all the unitary and anti-unitary symmetries. These results show that braiding statistics remains robust under real random hopping disorders. This highlights the crucial role of unitary symmetries in maintaining the robustness of SPNA statistics in bosonic SPT phases and the resilience of the system against real hopping disorders.

\section{Experimental schemes}\label{sec:experm}

\begin{figure}[t]
    \centering
    \includegraphics[scale=0.75]{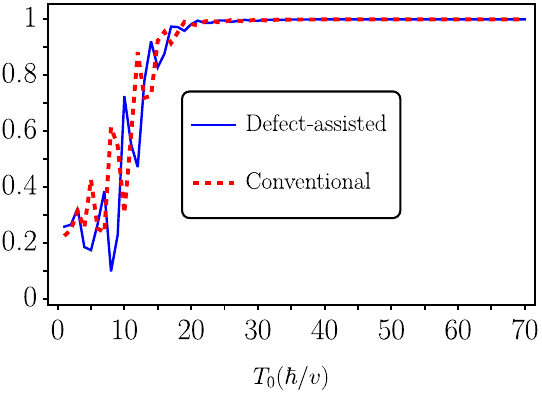}
    \caption{Numerical calculations of the average fidelity of the braiding matrix versus the time of elementary moving step $T_0$ in the experimental setup. The blue (red dotted) curve represents the average fidelity of the defect-assisted (conventional) braiding. Here we use a time-step precision of $\Delta t = 0.05$.}
    \label{fig:experAveFidelity}
\end{figure}

Finally, we demonstrate the high experimental feasibility of implementing SPNA statistics within the bosonic SPT phase using the state-of-the-art quantum simulators. To facilitate this, we consider a minimal configuration in the dimerized limit comprising four spins, as depicted in Fig.~\ref{fig:expconfig}(a). The rationale for selecting this minimal model is that the energy gap of the system is largely independent of its size. Initially, we set $v_1=v_3=0$ and prepare the system into the ground states. For the first (second) type of braiding scheme, the ground states are $\ket{\beta_1}=\frac{1}{\sqrt{2}}(\ket{1100}-\ket{1010})$ ($\ket{\beta_1}=\ket{1000}$) and $\ket{\beta_2}=\frac{1}{\sqrt{2}}(\ket{0101}-\ket{0011})$ ($\ket{\beta_2}=\ket{0001}$), as shown in Figs.~\ref{fig:expconfig}(b) and \ref{fig:expconfig}(c). The braiding of zero modes is achieved through the adiabatic tuning of spin-exchange couplings following the functions in Fig.~\ref{fig:expconfig}(d). The braiding in this minimal setup involves three elementary steps.

A quantum processor based on superconducting qubits is a promising platform for implementing our experimental scheme~\cite{Frank2019Quantum}. The tunable spin-exchange coupling in Eq.~\eqref{eq:expham} can be realized by employing couplers (additional qubits) between superconducting qubits [Fig.~\ref{fig:expconfig}(e)]. The coupler introduces a second-order process that adds a coupling channel on top of the direct qubit-qubit interaction~\cite{Yan2018,Xu2020}. The effective qubit-qubit coupling is
\be\label{eq:effhamexp}
\tilde{H}_{q-q}=\left[\frac{g_1 g_2}{2\Delta}+\frac{g_{12}}{2}\right]\left(\sigma_1^{x} \sigma_2^{x}+\sigma_1^{y} \sigma_2^{y}\right),
\ee
where parameter $\Delta = w_q-w_c$ denotes the detuning between the qubit and coupler and coupling strengths $g_1$, $g_2$, $g_{12}$ are fixed design parameters determined by the fabrication of the superconducting circuit. By modulating $\Delta$, the effective coupling strength can vary from zero to a significant magnitude. Recent technological advances in suppressing unwanted interactions~\cite{Chen2014Qubit,Ku2020} make this approach viable. A superconducting qubit tri-junction configuration in Fig.~\ref{fig:expconfig}(f) can be fabricated~\cite{Heras2014Digital,Salath2015}, with the braiding process executed by preparing initial states and adiabatically tuning coupling strengths as outlined in Fig.~\ref{fig:expconfig}(b). The quantum tomography method~\cite{Matthias2006}, which is commonly used in superconducting qubit systems, allows for precise measurement of the braiding results.

To determine the experimental parameters, we calculate the average fidelity as a function of the elementary moving step time $T_0$ [Fig.~\ref{fig:experAveFidelity}]. The fidelity calculations indicate that the dominate factor limiting the performance of the scheme is the adiabaticity of braiding process, further supporting the conclusion that the system is gap-protected. Moreover, the saturation of the fidelity at large $T_0$ further confirms the topological nature of both classes of braiding statistics, since dynamical phases would vary with time and thus could not produce a stable plateau. The two classes of statistics originate from non-Abelian Berry phases in distinct topological classes, as shown in Methods~\ref{sec:berry} and confirmed by numerical tests demonstrating robustness against local modifications of the evolution functions in Fig.~\ref{fig:expconfig}(d). From Fig.~\ref{fig:experAveFidelity}, We find that $\frac{T_0v}{\hbar}>25$ achieves average fidelities $F>0.99$, where $v$ represents the maximum spin-exchange coupling strength during the braiding. This implies that the total braiding time $T=3T_0>\frac{75\hbar}{v}$. To reduce the time required for the braiding operation, it is essential to make the spin-exchange coupling sufficiently large. With typical superconducting qubit coupling strengths reaching up to $20$ MHz~\cite{Salath2015}, the minimum braiding time is $3.75\mu s$ for a fidelity over $0.99$. The decoherence time $T_2$ for superconducting qubits is around $2\sim10\mu s$~\cite{Bylander2011}. We also note that implementing experiments with superconducting qubits by applying the non-adiabatic shortcut methods may further reduce the braiding time.

Our proposed scheme is not limited to superconducting qubit systems, but also applicable to other experimental systems, such as the Rydberg atom arrays. Recent developments in optical tweezer technology have enabled the manipulation of neutral atoms into the minimal tri-junction configurations with ~\cite{Dolev2022Atomarray,Daniel2016Assembler,Manuel2016}. Atoms can be prepared in the Rydberg state $nS_{\frac{1}{2}}$ and coupled to the $nP_{\frac{1}{2}}$ state using microwave fields~\cite{de2019observation}, with these levels $nS_{\frac{1}{2}}$ and $nP_{\frac{1}{2}}$ encoding spin states $\ket{0}$ and $\ket{1}$. The dipole-dipole interaction between the s- and p-levels of adjacent Rydberg atoms facilitates the spin-exchange coupling~\cite{Browaeys2016dipole}, with a typical interaction strength of approximately $10$ MHz~\cite{Saffman2010Rydberg,Antoine2020Many} by taking suitable distance between atoms. The tuning of coupling strength can be achieved by the laser-assisted dipole-dipole interactions scheme~\cite{Yang2022}. In this scheme, additional optical lights are applied to give an AC Stark shift to the Rydberg state at each site. By controlling the light field strength on at different sites, one can realize the effective Zeeman splitting offsets to tune the coupling strength. For sufficiently large principal quantum number $n$, Rydberg states have long enough lifetimes to complete the braiding process~\cite{beterov2009}. For our purpose, the braiding process must be finished within the decoherence time $T_2$, which is over $10 \mu s$ for typical Rydberg atom states with $n>60$~\cite{Bernien2017}. According to Fig.~\ref{fig:experAveFidelity}, a braiding process with fidelity $0.99$ takes over $7.5\mu s$, which is shorter than $T_2$, allowing the adequate time for measurements.

\section{Conclusions and Discussion}\label{sec:outlook}
In summary, we discover a new paradigm of SPNA statistics in a representative bosonic SPT phase protected by unitary symmetry, with a systematic and rigorous theory being established. The topological zero modes exhibit two distinct classes of braiding statistics: the conventional and defect-assisted unconventional braiding statistics, the latter being a fractionization of the former and manifesting a novel nonlinear transformation. The braiding statistics can be extended to generic bosonic SPT phases with real Hamiltonian, representing a fundamental dynamical character of such topological phases.
We propose feasible tri-junctions with or without the assistance of a controlled defect to realize the two classes of braiding statistics in the real systems, with which 
the non-commutative quantum gates are achieved in the single-qubit setting formed by two zero modes. The results also facilitate the realization of various single- and two-qubit quantum gates. Experimental schemes are proposed and discussed in detail for observation of these predictions with high-fidelity, paving the way for the experimental study in the near future.

Pertinent directions for future investigation include the SPNA statistics of topological zero modes in the more generic bosonic SPT phases, including those with complex Hamiltonians. An interesting question is whether the phase gate, which is essential for universal quantum computing, can be realized via braiding of zero modes in bosonic SPT phases. This presents an exciting avenue for the future research.

We note that this work holds the great potential to promote the first experimental realization of non-Abelian quasiparticles and their braiding statistics in quantum many-body phases. Currently, the rigorous verification of MZMs and their non-Abelian braiding in topological superconductors faces substantial challenges. Digital simulations of non-Abelian braiding are actively pursued on quantum computing platforms~\cite{google2023non,xu2023digital,iqbal2024non}, whereas realizing non-Abelian quasiparticles and their braiding statistics poses a fundamentally distinct and paramount task in topological quantum matter. The present study, with its full experimental feasibility, paves the way for the ultimate experimental observation of non-Abelian quasiparticles in topological quantum phases and their braiding statistics in the near future, which may trigger the SPT quantum computation, as a new route for topological quantum computation.


\begin{figure}[t]
\centering
\includegraphics[scale=0.22]{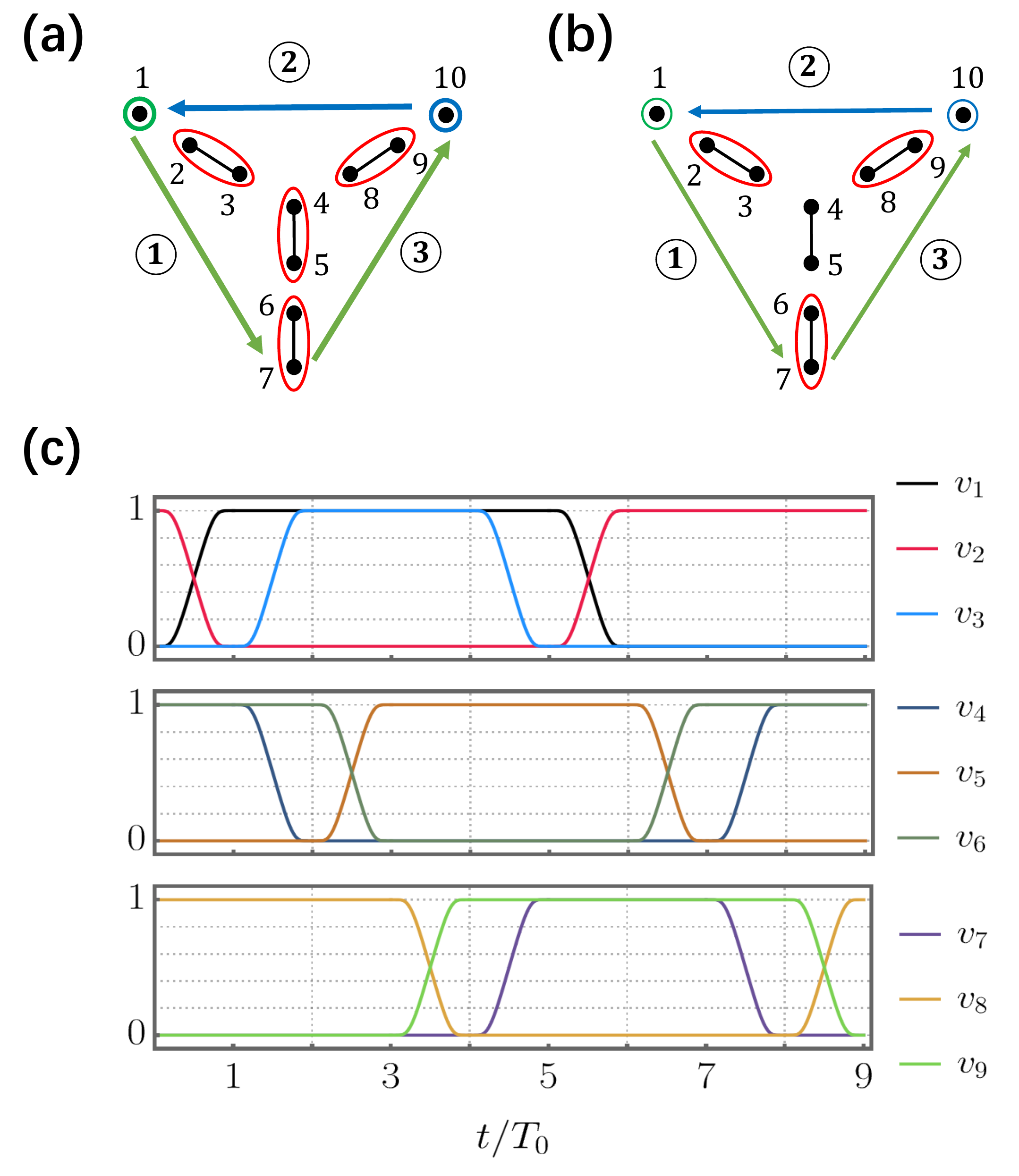}
\caption{Tri-junctions in the dimerized limit with $10$ sites, where site $1$ ($10$) hosts the left (right) zero mode. The ground state $\ket{\beta_1}$ ($\ket{\beta_2}$) corresponds to site $1$ ($10$) being occupied. The green and blue arrows indicate  the direction in which the zero mode moves during braiding. (a) The tri-junction that realizes the conventional braiding. (b) The tri-junction that realizes the defect-assisted braiding, with sites $4$ and $5$ serving as the local defect. (c) Time dependence of coupling strengths in the dimerized tri-junction for braiding realization. The horizontal axis denotes the number of elementary steps.}
    \label{fig:zeromode}
\end{figure}

\section{Methods}
\subsection{The floquet theory for effective Hamiltonian}\label{sec:Floquet}
In this section, we derive the effective Hamiltonian Eq.~\eqref{eq:dbana} from the periodically driven Hamiltonian
\begin{equation}\label{eq:periodHam}
H(t)=H_0+2 H'_1 \cos \omega t+2 H'_2 \sin \omega t .
\end{equation}
The effective Hamiltonian obtained by Floquet theory up to the order of $1/\omega^2$ is
\begin{eqnarray}\label{eq:floqueteffham}
H_E &=& H_0 + \frac{1}{2\omega^2}([V_1, H_0], V_{-1}] + \text{ h.c.})
\nonumber\\
&&+\frac{1}{\omega}\left[V_1, V_{-1}\right]+\mathcal{O}\left(\frac{1}{\omega^3}\right) \\
&=& H_0+\frac{i}{\omega}\left[H'_1, H'_2\right]+\frac{1}{\omega^2}\left[2\left(H'_1 H_0 H'_1+H'_2 H_0 H'_2\right)\right. \nonumber\\
&&\left.+\left((H'_1)^2 H_0+(H'_2)^2 H_0+\text { h.c. }\right)\right]+\mathcal{O}\left(\frac{1}{\omega^3}\right), \nonumber
\end{eqnarray}
where $V_1=H'_1-i H'_2$ and $V_{-1}=V_1^{\dagger}$. Substituting
\begin{eqnarray}
    H'_1&=&\delta_{1}(b_{1a}^{\dagger}b_{1b}+b_{1b}^{\dagger}b_{1a}), \nonumber
    \\
    H'_2&=&\delta_{2}(ib_{1a}^{\dagger}b_{1b}-ib_{1b}^{\dagger}b_{1a}), \nonumber
\end{eqnarray}
into Eq.~\eqref{eq:floqueteffham} gives rise to the Eq.~\eqref{eq:dbana}.

\subsection{Braiding operators from the Yang-Baxter equation}\label{sec:yangbaxter}
In this section, we derive the braiding operators by solving the Yang-Baxter equation, thereby establishing the SPNA statistics for hard-core bosonic zero modes on a mathematically rigorous footing. We focus on the case protected by the unitary symmetries of particle-hole transformation and particle-number conservation. For two zero modes, the most general effective Hamiltonian governing the braiding process is:
\begin{equation}\label{eq:effective}
H_{E} = \lambda G_{12}(\alpha) + \delta P_{12},
\end{equation}
where $\lambda$, $\delta \in \mathbb{R}$, $G_{12}(\alpha) \equiv e^{i\alpha}\beta_1^{\dagger}\beta_2 + e^{-i\alpha}\beta_2^{\dagger}\beta_1$ with $\alpha \in [0,2\pi)$, and $P_{12} \equiv G_{12}^2 = n_1 +n_2 -2n_1 n_2$ is the projector onto the half-filled subspace $\mathcal{H}_0$. These operators satisfy
\begin{equation}
G^3=GP=PG = G, \quad P^2 = P.
\end{equation}
The braiding operator takes the form
\begin{eqnarray}\label{eq:brop}
    U_{12} &=& e^{-iH_{E}T} = e^{-i\lambda T G_{12}(\alpha)}e^{-i\delta T P_{12}}
    \\
    &=& I -ie^{-i\theta}\text{sin}(\varphi)G_{12}(\alpha) + (e^{-i\theta}\text{cos}(\varphi)-1)P_{12}, \nonumber
\end{eqnarray}
where the subscript indicates the zero-mode pair being exchanged, and we define $\lambda T \equiv \varphi$ and $\delta T \equiv \theta$ for brevity.

When more than two zero modes are present, the Yang-Baxter equation~\cite{Yang1967,Baxter1972} ensures consistency between two different sequences of braiding three zero modes:
\be\label{eq:YBeq}
U_{12}U_{23}U_{12} = U_{23}U_{12}U_{23}.
\ee
substituting Eq.~\eqref{eq:brop} into Eq.~\eqref{eq:YBeq} gives:
\begin{widetext}
\begin{eqnarray}\label{eq:ybexpre}
&&a G_{12} + 2 a b G_{12} - a G_{23} - 2 a b G_{23} + a^2 P_{12} + b P_{12} + b^2 P_{12}
- a^2 P_{23} - b P_{23}  +
 a b^2 P_{12} G_{23} P_{12} + a^2 b P_{12} G_{23} G_{12}  \nonumber \\ &&- b^2 P_{23} + a^3 G_{12} G_{23} G_{12} +
 a^2 b G_{12} G_{23} P_{12} + a^2 b G_{12}P_{23}G_{12}  +
 a b^2 G_{12} P_{23} - a b^2  P_{12} G_{23} + a b^2  P_{23} G_{12}
  \nonumber \\ &&- a^3 G_{23}G_{12}G_{23}  -
 a^2 b G_{23} G_{12} P_{23} - a^2 b G_{23} P_{12} G_{23} - a b^2 G_{23} P_{12} - a b^2 P_{23} G_{12} P_{23} - a^2 b P_{23} G_{12} G_{23}  = 0,
 \end{eqnarray}
 \end{widetext}
where we have defined $a \equiv - ie^{-i\theta}\text{sin}(\varphi)$ and $b \equiv e^{-i\theta}\text{cos}(\varphi)-1$, and used $[P_{12},P_{23}]=0$ to simplify intermediate steps. Using the additional algebraic relations
\begin{eqnarray}
  \{P_{12},G_{23}(\alpha)\} &=& G_{23}(\alpha), \\
  \{P_{23},G_{12}(\alpha)\} &=& G_{12}(\alpha),
\end{eqnarray}
the above expression Eq.~\eqref{eq:ybexpre} reduces to
\begin{eqnarray}
    (a+2ab &+& ab^2) (G_{12}(\alpha) - G_{23}(\alpha))
    \\
     &+& (a^2+b+b^2+a^2b)(P_{12}-P_{23}) = 0. \nonumber
\end{eqnarray}
Thus, $a$ and $b$ must satisfy
\begin{equation} \label{eq:ybereduce}
    a+2ab+ab^2 = 0, \quad a^2+b+b^2+a^2b = 0.
\end{equation}
For simplicity and without loss of physical content, we restrict $\theta$ and $\varphi$ to the range $(-\pi, \pi]$; more general solutions are $\theta+2n\pi$ and $\varphi+2m\pi$ with $m,n\in \mathbb{Z}$. The nontrivial solutions of Eq.~\eqref{eq:ybereduce} are $\varphi = \pm \frac{\pi}{2}$. The choice of the plus or minus sign is merely a convention, as the two cases are related by Hermitian conjugation accompanied by reversing the sign of $\theta$. Physically, the sign corresponds to the braiding direction—clockwise or counterclockwise. In this work, we adopt the $\varphi = -\frac{\pi}{2}$ convention, and the corresponding braiding operator is
\begin{equation}\label{eq:solution}
    U_{12} = I + ie^{-i\theta}G_{12}(\alpha) - P_{12}.
\end{equation}

The solution of the braiding operator in Eq. \eqref{eq:solution} to the Yang-Baxter equation depends on a continuous parameter $\theta$. This dependence originates from the term $\propto P_{12}$ in $H_E$, which physically describes an interaction between the two zero modes: $P_{12}=1$ when exactly one zero mode is occupied and $P_{12}=0$ otherwise. Such an interaction leads to phase accumulation during braiding, thereby producing a continuous family of solutions for the braiding operator. The fact that the Hamiltonian is real imposes an additional constraint: the Berry phase must be real. Consequently, for a full braiding process, the corresponding transformation is
\begin{equation}\label{eq:fullbr}
U^2_{12}= 1-(1+e^{-2i\theta})P_{12} = \pm I.
\end{equation}
We refer to the cases $U^2 = I$ and $U^2 = -I$ as conventional braiding statistics and defect-assisted braiding statistics, respectively. When both zero modes are either occupied or empty, $P_{12}=0$, yielding $U^2_{12}=1$. This result is physically consistent because the bosonic commutation relations of the zero-mode operators require that the singlet sectors $\ket{0,0}$ and $\ket{1,1}$ are invariant under braiding. Nevertheless, when exactly one zero mode is occupied, we have $U_{12}^2 = -e^{-2i\theta}$. This condition leads to two distinct classes of solutions for the parameter $\theta$: (i) $\theta = \pm \pi/2$, and (ii) $\theta = 0$ or $\pi$, corresponding to two classes of effective Hamiltonians. {Moreover, as shown in Fig.~\ref{fig:trijuncwave} and in Methods~\ref{sec:berry}, the braiding transformation of the ground states is real. This requires that the coefficients in Eq.~\eqref{eq:solution} be real, which in turn fixes $\alpha$ with a given $\theta$.} For the firt class, the effective Hamiltonian realizes the conventional braiding statistics, with the braiding operator
\begin{eqnarray}\label{eq:convbap}
    U^{\text{conv}}_{12} &=& I + G_{12}(0) - P_{12}, \nonumber
    \\
    &=& I+ \beta_1^{\dagger}\beta_2+ \beta_2^{\dagger}\beta_1 -n_1-n_2+2n_1n_2,
\end{eqnarray}
where we take $\theta = \pi/2$ by convention, as the opposite choice only introduces an overall $-1$ factor to the braiding matrix in the subspace $\mathcal{H_0}$. {Since the coefficients in Eq.~\eqref{eq:convbap} are already real, the $\alpha$ takes the value $0$ (Also, $\alpha=\pi$ only contributes an overall factor of $-1$).}

For the second class, we choose $\theta = 0$ with the associated $\alpha=\frac{\pi}{2}$ to ensure that the braiding operator is real. It then reads
\begin{eqnarray}\label{eq:defbap}
U^{\text{def}}_{12} &=& I + iG_{12}(\frac{\pi}{2}) - P_{12}, \nonumber
\\
&=& I -\beta_1^{\dagger}\beta_2 + \beta_2^{\dagger}\beta_1  -n_1-n_2+2n_1n_2,
\end{eqnarray}
where the opposite choice differs only by Hermitian conjugation. These results, Eqs.~\eqref{eq:convbap} and \eqref{eq:defbap}, are confirmed with the numerical results and the exact solution of the wavefunction evolution during the braiding process, in which all the coefficients are real, as shown in the following Methods~\ref{sec:berry}.





\subsection{The non-Abelian Berry phase and topological classification}\label{sec:berry}

In this section, we compute the non-Abelian Berry phase associated with the braiding process using the Wilczek-Zee connection. We show that conventional and defect-assisted braiding statistics belong to distinct topological classes and provide a geometric interpretation for both classes of non-Abelian statistics.

The Wilczek-Zee connection generalizes Berry's geometric phase to the case of degenerate states~\cite{Wilczek1984}. Consider a Hamiltonian $H(\vec{R})$ with a degenerate subspace of energy $E_m$, spanned by orthonormal basis $\{\ket{\psi_{m,a}}\}$ with $a=1,...,d_m$ denoting the multiplicity, and depending on time through parameters $\vec R(t)$. Under a time-dependent evolution $\vec R(t)$, the state evolves as
\begin{equation}
\ket{\Psi(t)} = \sum_{m,a} e^{-i\int E_{m}(t)dt} c_{m,a}(t) \ket{\psi_{m,a}(t)},
\end{equation}
where $\ket{\psi_{m,a}(t)}$ are instantaneous orthonormal eigenstates $H(\vec R(t))\ket{\psi_{m,a}(t)}=E_m(t)\ket{\psi_{m,a}(t)}$. The Schrödinger equation then gives
\begin{widetext}
\begin{eqnarray}\label{eq:adaeq}
    \dot{c}_{n,b}(t) &=& -\sum_{a}\langle \psi_{n,b}(t)\vert \dot{\psi}_{n,a}(t) \rangle c_{n,a}(t) - \sum_{m\neq n, a}e^{-i\int(E_m-E_n)dt}\frac{\langle \psi_{n,b}(t)\vert \dot{H}(t) \vert \psi_{m,a} \rangle}{E_m-E_n} c_{m,a}(t), \nonumber
    \\
    &\approx& -\sum_{a}\langle \psi_{n,b}(t)\vert \dot{\psi}_{n,a}(t) \rangle c_{n,a}(t)
    =: -\sum_{a}A^n_{ba}(t) c_{n,a}(t),
\end{eqnarray}
\end{widetext}
where we use the adiabatic approximation
\[
\vert \frac{\langle \psi_{n,b}(t)\vert \dot{H}(t) \vert \psi_{m,a} \rangle}{E_m-E_n} \vert \ll \vert A^n_{ba} \vert,
\]
to neglect the second term. Here, $A^n_{ba}\equiv\langle \psi_{n,b}(t)\vert \dot{\psi}_{n,a}(t) \rangle$ is the Wilczek-Zee connection. The formal solution of Eq.~\eqref{eq:adaeq} is
\begin{equation}
c_{n,b}(t) = \sum_{a} \mathcal{P}e^{-\int A^n_{ba}(t) dt} c_{n,a}(0),
\end{equation}
where $\mathcal{P}$ denotes path ordering.

For simplicity, we focus on the minimal braiding setups illustrated in Fig.~\ref{fig:expconfig}(b) and Fig.~\ref{fig:expconfig}(c). The extension to larger tri-junction configurations is straightforward. The Hamiltonian of the minimal setup is
\begin{equation}
   H(\vec R(t)) = v_1(t)b_1^{\dagger}b_2 + v_2(t)b_2^{\dagger}b_3 + v_3(t)b_2^{\dagger}b_4 + h.c.,
\end{equation}
where the parameter vector is defined as $\vec R = (v_1, v_2, v_3)^T$, with $v_1(t)$, $v_2(t)$, and $v_3(t)$ time-dependent coupling parameters, e.g., illustrated in Fig.~\ref{fig:expconfig}(d). The parameter space consists of three patches joined along their edges, forming a conical surface with a singular point at $\vec R = (0,0,0)^T$ where the gap closes,
\begin{equation}\label{eq:paraspace}
\vec R^{T} \in ([0,1],0,[0,1]) \cup (0,[0,1],[0,1]) \cup ([0,1],[0,1],0).
\end{equation}
The braiding process corresponds to a closed loop encircling this singularity, which gives rise to a non-Abelian Berry phase. Because the Hamiltonian is real, one can always choose a real instantaneous basis at each time through an appropriate gauge choice. Consequently, the Wilczek-Zee connection is purely real and satisfies
\begin{equation}
A^n_{aa} = 0, \qquad A^n_{ab} = -A^n_{ba}.
\end{equation}

We first consider defect-assisted braiding. In this case, the system is occupied by a single hard-core boson, and at $t=0$ the zero-energy subspace (the ground-state subspace in the presence of a local defect) is spanned by the basis states $(\ket{1}, \ket{4})$, where $\ket{j} \equiv b_j^{\dagger}\ket{\Uparrow}$ with $\ket{\Uparrow}$ denoting the state in which all spins are in the up state. To ensure smooth evolution of the basis vectors, we adopt the following gauge choice:
\begin{widetext}
\begin{equation}\label{eq:defgauge}
  \begin{cases}
   \left(\frac{1}{\sqrt{v(t)^2+u(t)^2}}(v(t)\ket 1 - u(t)\ket 3), \ket 4\right), \quad  0 \leq t < T_0 \\
  \left(-\ket 3, \frac{1}{\sqrt{v(t-T_0)^2+u(t-T_0)^2}}(v(t-T_0)\ket 4-u(t-T_0)\ket 1)\right), \quad    T_0 \leq t < 2T_0\\
  \left(\frac{1}{\sqrt{v(t-2T_0)^2+u(t-2T_0)^2}}(-v(t-2T_0)\ket 3 + u(t-2T_0)\ket 4)), -\ket 1\right), \quad   2T_0 \leq t \leq 3T_0
\end{cases}
\end{equation}
\end{widetext}
where $u(t)$ and $v(t)$ are smooth interpolation functions satisfying the boundary conditions $u(0)=0$, $u(T_0)=1$, $v(0)=1$ and $v(T_0)=0$. For conventional braiding, the system is occupied by two hard-core bosons. At $t=0$, the ground-state subspace is spanned by the basis $\left(\tfrac{1}{\sqrt{2}}(\ket{12}-\ket{13}), \tfrac{1}{\sqrt{2}}(\ket{24}-\ket{34})\right)$, where $\ket{ij}=b_i^{\dagger}b_j^{\dagger}\ket{\Uparrow}$. With an appropriate gauge choice, the basis states evolve smoothly as
\begin{widetext}
{\footnotesize
\begin{equation}\label{eq:convgauge}
  \begin{cases}
   \frac{1}{\sqrt{2(u(t)^2+v(t)^2)}}\left(v(t)\ket{12} + u(t)\ket{23} -\sqrt{u(t)^2+v(t)^2}\ket{13}, -v(t)\ket{34}-u(t)\ket{14}+\sqrt{u(t)^2+v(t)^2}\ket{24}\right),  0 \leq t < T_0 \\
  \frac{1}{\sqrt{2(u(t')^2+v(t')^2)}}\left(-v(t')\ket{13} - u(t')\ket{34} +\sqrt{u(t')^2+v(t')^2}\ket{23}, v(t')\ket{24}+u(t')\ket{12}-\sqrt{u(t')^2+v(t')^2}\ket{14}\right),  T_0 \leq t < 2T_0\\
 \frac{1}{\sqrt{2(u(t'')^2+v(t'')^2)}}\left(v(t'')\ket{23} + u(t'')\ket{24} -\sqrt{u(t'')^2+v(t'')^2}\ket{34}, -v(t'')\ket{14}-u(t'')\ket{13}+\sqrt{u(t'')^2+v(t'')^2}\ket{12}\right),   2T_0 \leq t \leq 3T_0
\end{cases}
\end{equation}
}
\end{widetext}
where $t'\equiv t-T_0$ and $t''\equiv t-2T_0$. Eq.~\eqref{eq:defgauge} and Eq.~\eqref{eq:convgauge} shows that the basis states are changed at the end of the braiding. To compute the non-Abelian Berry phase, we adiabatically return the basis to its initial configuration using the following gauge transformations for the time interval $2T_0\leq t< 3T_0$:
\begin{eqnarray}
  &&\text{defec-assisted}: \quad g^{\text{def}} = e^{i\phi(t)\sigma^y},\\
    &&\text{convetional}: \quad g^{\text{conv}} =  \sigma^z e^{i\phi(t)\sigma^y}.
\end{eqnarray}
Here, $\phi(t)$ is a smooth functions satisfying $\phi(2T_0)=0$ and $\phi(3T_0)=\frac{\pi}{2}$. These gauge transformations act on the basis from the right, i.e., $(\ket{\psi_1}, \ket{\psi_2})\cdot g$. The Wilczek-Zee connection is then given by
\begin{equation}
    A = i\sigma^y\dot \phi(t),
\end{equation}
 for both defect-assisted braiding and conventional braiding. The corresponding non-Abelian Berry phase is
\begin{eqnarray}
    U^{\text{def}}&=&\mathcal{P}e^{-\int_{2T_0}^{3T_0}A dt} = -i\sigma^y \in SO(2),
    \\
    U^{\text{conv}}&=&\mathcal{P}e^{-\int_{2T_0}^{3T_0}Adt} \sigma^z  = \sigma^x \in O(2).
\end{eqnarray}
The key difference between $U^{\text{def}}$ and $U^{\text{conv}}$ is whether the holonomy involves the reflection operator $\sigma^z$. In conventional braiding, this reflection acts as a branch cut: crossing it flips the orientation of the basis vectors, so the subsequent phase accumulation changes sign.

Mathematically, the determinants of two braiding matrices $\text{det}(U^{\text{def}})=1$ and $\text{det}(U^{\text{conv}})=-1$, arising from the fact that a real system only allows phase factors $\pm 1$, correspond to two topological invariants associated with the first Stiefel-Whitney class of a real vector bundle. More precisely, the evolution of the two-dimensional zero-mode subspace defines a rank-$2$ real vector bundle $E$ over the base manifold $S^1$, which corresponds to a closed loop in parameter space. Since the basis states are real and normalized, each fiber is a circle $S^1$, so $E$ is a circle bundle over a circle. Such bundles fall into two distinct classes: the orientable torus and the non-orientable Klein bottle. Their classification is given by the first Stiefel-Whitney class, defined as
\begin{equation}
    w_1(E) := \frac{1+\text{det}(U)}{2} \in H^1(S^1;\mathbb{Z}_2)\cong \mathbb{Z}_2.
\end{equation}
The class $w_1(E)=1$ corresponds to an orientable bundle, where the structure group reduces from $O(2)$ to $SO(2)$, matching the defect-assisted braiding. The class $w_1(E)=0$ corresponds to a non-orientable bundle, where the structure group remains $O(2)$, as in the conventional braiding. We summarize the correspondences in Table.~\ref{tab:corr}. This provides a topological interpretation of the two classes of braiding statistics.

\begin{table}[h]
\centering
\begin{tabular}{cccc}
\hline\hline
\small the first &  topological & braiding & structure \\ \small Stiefel-Whitney class  & invariant & matrix & group \\
\hline
$w_1(E)=1$ & $\text{det}(U^{\text{def}})=1$ & $U^{\text{def}}=-i\sigma^y$ & SO(2) \\
$w_1(E)=0$  & $\text{det}(U^{\text{conv}})=-1$ & $U^{\text{conv}}=\sigma^x$ & O(2) \\
\hline\hline
\end{tabular}
\caption{Correspondences between the first Stiefel-Whitney class, the associated topological invariants, the two classes of braiding statistics, and structure groups.}
\label{tab:corr}
\end{table}

\subsection{Average fidelity and unitarity}\label{sec:fidelity}

We introduce average fidelity as a quantitative measure to evaluate the discrepancy between actual and ideal braiding results~\cite{Line2007}. Given an arbitrary initial state $\ket{\psi}$, the fidelity of an operator $U$ can be characterized by
\begin{equation}\label{eq:fidelity0}
f(U,O)=|\bra{\psi}O^{\dagger}U\ket{\psi}|^2,
\end{equation}
where $O\ket{\psi}$ defines the final state after an ideal operation. Then the average fidelity is defined as the fidelity averaged over the subspace encoding the qubit: $F(U,O) \equiv \int_{S^3} |\langle \psi|O^{\dagger}U |\psi \rangle |$, which yields
\begin{eqnarray}\label{eq:avefidelity1}
F(U,O)=\frac{1}{6} (\text{Tr}(U^{\dagger}U)+|\text{Tr}(O^{\dagger}U)|^2).
\end{eqnarray}
This quantity characterizes the agreement between the outcomes of the operator $U$ and the ideal operator $O$ when acting on states within the qubit subspace. For our study, the ideal braiding matrix $O$ is either $i\sigma_y$ for the defect-assisted unconventional braiding or $\sigma^z$ for the conventional braiding in the basis $\{\ket{\eta},\ket{\xi}\}$. If the operator $U$ is unitary, the average fidelity has a lower bound
\begin{eqnarray}\label{eq:avefidelity2}
F_{\rm min}(U,O)=\frac{1}{3}.
\end{eqnarray}
To quantify how unitary the braiding matrix \( U \) is, we further define the following quantity:
\begin{eqnarray}\label{eq:unitarity3}
\Tilde{U}(U) &\equiv& \int_{S^3} |\langle \psi|U^{\dagger}U |\psi \rangle | \nonumber
\\
&=& \frac{1}{6} (\text{Tr}(U^{\dagger}UU^{\dagger}U)+|\text{Tr}(U^{\dagger}U)|^2).
\end{eqnarray}
The braiding matrix \( U \) is unitary if and only if \(\Tilde{U}(U) = 1\).

\section*{}

\textbf{Data availability.}--The data that support the findings of this study are available from the corresponding author upon reasonable request.

\textbf{Code availability.}--All numerical codes in this paper are available upon request to the
authors.

\nocite{*}

%

\section*{}

\textbf{Acknowledgments.}--This work was supported by National Key Research and Development Program of China (2021YFA1400900), National Natural Science Foundation of China (No.~12425401 and No.~12261160368), Quantum Science and Technology-National Science and Technology Major Project (2021ZD0302000), and Shanghai Municipal Science and Technology Major Project (No.~2019SHZDZX01).

\textbf{Competing interests.}--The authors declare no competing interests.

\clearpage
\onecolumngrid
\section*{Supplementary Material}

\setcounter{section}{0}
\renewcommand{\thesection}{S\arabic{section}}
\setcounter{equation}{0}
\renewcommand{\theequation}{S\arabic{equation}}
\setcounter{figure}{0}
\renewcommand{\thefigure}{S\arabic{figure}}
\setcounter{table}{0}
\renewcommand{\thetable}{S\arabic{table}}

\section{CALCULATIONS IN THE DIMERIZED LIMIT}\label{sec:brdimer}

In this section, we study the two classes of braiding statistics in the dimerized limit, which provides a more intuitive understanding of the non-trivial geometric phases. We focus on specific dimerized tri-junctions that realize the first [Fig.~\ref{fig:zeromode}(a)] and the second [Fig.~\ref{fig:zeromode}(b)] types of braiding scheme.

We first compute the geometric phase of an elementary moving step illustrated in Fig.~\ref{fig:specandbraiding}(e). The dynamics of an elementary moving step is governed by the Hamiltonian $H_{\text{e}}$ in Eq.~\eqref{eq:emp}. In the dimerized limit, the parameters at $t=0$ are initialized with $u(0)=0$, $v(0)=v$. Adiabatically tuning $u(t)$ to $u(T_{0})=v$ and $v(t)$ to $v(T_{0})=0$ moves the zero mode from site $1$ to site $3$. The geometric phase obtained by the zero mode depends on the particle-number occupation on sites $2$ and $3$. In both the single-particle subspace spanned by $\{b_{1}^{\dagger}|\Uparrow\rangle,b_{2}^{\dagger}|\Uparrow\rangle,b_{3}^{\dagger}|\Uparrow\rangle\}$
and the two-particle subspace spanned by $\{b_{2}^{\dagger}b_{3}^{\dagger}|\Uparrow\rangle,b_{3}^{\dagger}b_{1}^{\dagger}|\Uparrow\rangle,b_{1}^{\dagger}b_{2}^{\dagger}|\Uparrow\rangle\}$, where $\ket{ \Uparrow}$ denotes the state with all spins up,
the $H_{\text{e}}$ is
\begin{equation}
H_{\text{e}}=\begin{pmatrix}0 & u(t) & 0\\
u(t) & 0 & v(t)\\
0 & v(t) & 0
\end{pmatrix}.
\end{equation}
The initial states are eigenstates of $H_{\text{e}}$ with site $1$ occupied. In the single-particle subspace, the initial state is the zero-energy
state $\left[v(t)b_{1}^{\dagger}-u(t)b_{3}^\dagger\right]|\Uparrow\rangle$. In the two-particle subspace, the initial
state is the ground state $\left[v(t)b_{1}^\dagger b_{2}^\dagger+u(t)b_{2}^\dagger b_{3}^\dagger - \sqrt{u(t)^2+v(t)^2} b_3^\dagger b_1^\dagger\right]|\Uparrow\rangle$.
Then after the elementary moving step $u(t): 0 \to v$, $v(t): v \to 0$,
these states evolve as follows:
\begin{eqnarray}
b_{1}^{\dagger}|\Uparrow\rangle & \to & -b_{3}^{\dagger}|\Uparrow\rangle,\\
b_{1}^{\dagger}(b_{2}^{\dagger}-b_{3}^{\dagger})|\Uparrow\rangle & \to & b_{3}^{\dagger}(b_{2}^{\dagger}-b_{1}^{\dagger})|\Uparrow\rangle.
\end{eqnarray}
In the single-particle subspace, the zero mode changes its position and acquires a $\pi$ phase. In the two-particle subspace, the zero mode simply changes its position without acquiring an additional phase.

We now consider the specific dimerized tri-junctions depicted in Figs.~\ref{fig:zeromode}(a) and \ref{fig:zeromode}(b). The braiding process involves $9$ elementary moving steps with a total braiding time $T=9T_0$. In the ground state of the dimerized tri-junction, hard-core bosons form dimer bond states (i.e., Bell states in the spin representation) between pairs of sites. In Fig.~\ref{fig:zeromode}(a), all sites except for edge sites $1$ and $10$ are occupied by dimers. In Fig.~\ref{fig:zeromode}(b), the pair sites $(4,5)$ are not occupied by a dimer and play the role of a local defect. During the braiding process, the filling on $(4,5)$ is crucial to the net geometric phase because it is the only pair of sites through which one zero mode passes once while another passes twice. The other pairs are either passed through once by both zero modes or twice by a single zero mode and thus do not contribute to the net geometric phase. In the case of Fig.~\ref{fig:zeromode}(a), since the local defect $(4,5)$ is half-filled, ground states acquire no geometric phase difference. Therefore, states $\ket{\beta_1}$ and $\ket{\beta_2}$ transform as Eq.~\eqref{eq:statetranstr}. In the case of Fig.~\ref{fig:zeromode}(b), however, the local defect $(4,5)$ is empty, ground states $\ket{\beta_1}$ and $\ket{\beta_2}$ acquire a $\pi$ geometric phase difference. Accordingly, states $\ket{\beta_1}$ and $\ket{\beta_2}$ transform as Eq.~\eqref{eq:statetransntr}.

\begin{figure*}[t]
    \centering
    \includegraphics[scale=0.25]{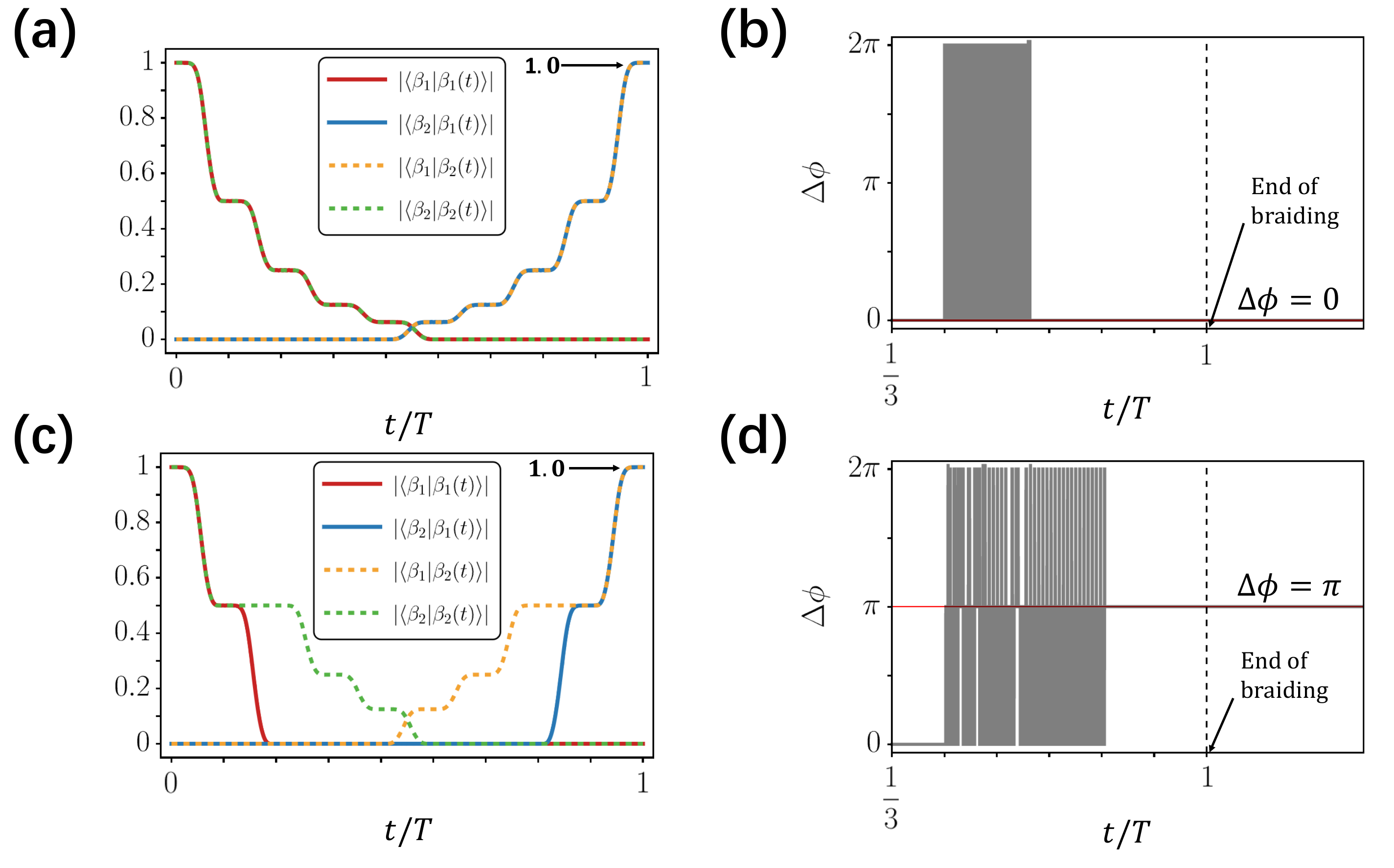}
    \caption{Numerical time evolution of zero-mode wavefunctions for two braiding schemes in the dimerized limit. (a) Amplitudes of and (b) phase difference between $\ket{\beta_1(t)}$ and $\ket{\beta_2(t)}$ for the first type of braiding scheme (conventional braiding). (c) Amplitudes and (d) phase difference for the second type of braiding scheme (defect-assisted braiding). The horizontal axis represents the evolution time in units of the total braiding time $T$. The strong and weak couplings take $v_{\text{max}}=1$ and $v_{\text{min}}=0$. Simulations are performed with a time step $\Delta t = 0.1$ and an elementary step time $T_0 = 80$. Each braiding involves $9$ elementary steps of zero-mode motion, yielding a total braiding time $T = 9T_0 = 720$. The red horizontal line indicates the theoretical prediction of $\Delta \phi = \text{Arg}(\langle \beta_2 | \beta_1(t) \rangle ) - \text{Arg}(\langle \beta_1 | \beta_2(t) \rangle) \, \text{mod} \, 2\pi$ for both schemes. The vertical dashed line marks the completion of the braiding process, after which the system is held static to show the stabilized value of $\Delta \phi$. The numerical results of $\Delta \phi$ for both classes of statistics are indicated in the figure.}
    \label{fig:wavefunction}
\end{figure*}

We numerically verify the above theoretical results. The time-dependent Hamiltonian used to implement both braiding schemes with tri-junctions is
\begin{equation}\label{eq:hamevo}
H(t) = \sum_{i=1,2,...,6,8,9}v_{i}(t)b^{\dagger}_{i+1}b_{i}+v_7(t)b^{\dagger}_{8}b_{4}+h.c..
\end{equation}
In the braiding we adjust the coupling strengths over time following the functions depicted in Fig.~\ref{fig:zeromode}(c). The rising (falling) slope in Fig.~\ref{fig:zeromode}(c) follow the form $\chi(t/T_0)$ ($1-\chi(t/T_0)$) defined in Eq.~\eqref{eq:timeevofunc}. The numerical results, shown in Fig.~\ref{fig:wavefunction}, match the theoretical predictions.

\section{ERROR PREVENTION THROUGH LOCAL DEFECT ISOLATION}\label{sec:effError}

\begin{figure}[h]
    \centering
    \includegraphics[scale=0.25]{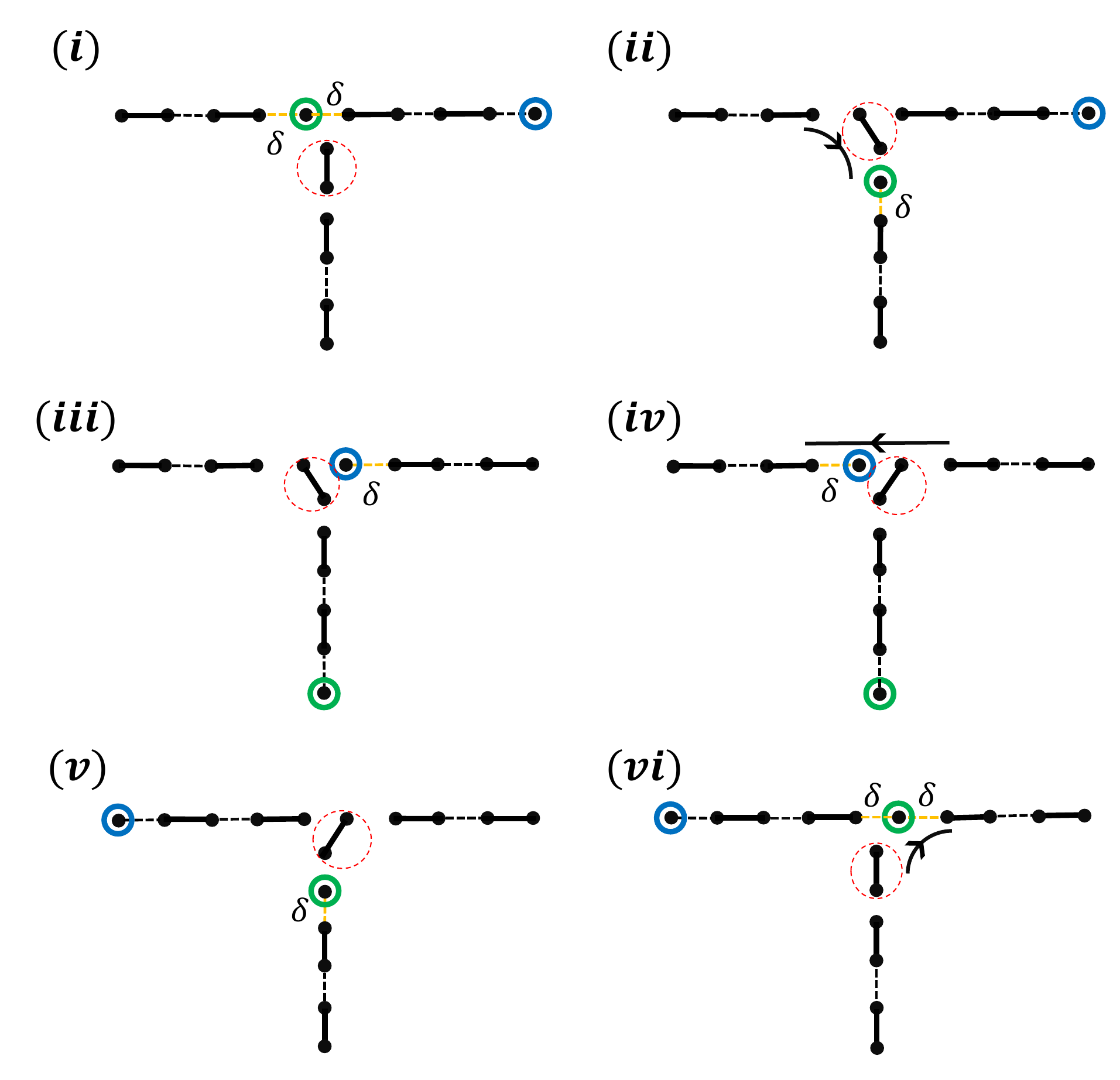}
    \caption{Imperfect defect-assisted braiding procedure. The golden dashed lines represent the residual coupling strength $\delta$ between the zero modes and other parts of the tri-junction as the zero modes pass through the local defect.}
    \label{fig:PassProcess}
\end{figure}

\begin{figure}[h]
    \centering
    \includegraphics[scale=0.65]{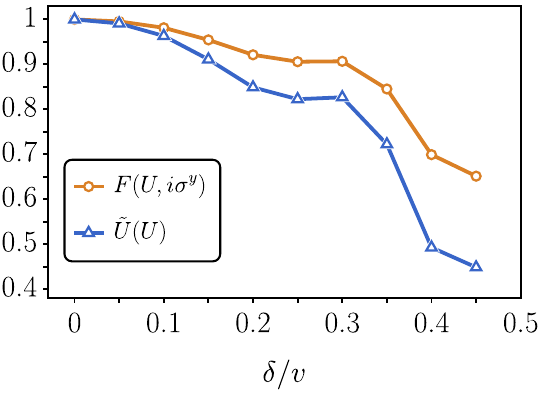}
    \caption{Numerical calculations of the average fidelity $F$ and unitarity $\tilde U$ as functions of the residual coupling strength $\delta$ (in the unit of maximal coupling strength $v$) for imperfect defect-assisted braiding procedures. }
    \label{fig:defctst}
\end{figure}

Defect-assisted braiding requires isolating the local defect to ensure that it returns to the original state after braiding. Failure to meet this condition causes errors in the braiding results. In the following, we analyze these errors and verify that our scheme through isolating the local defect effectively prevents them. Specifically, we examine an imperfect defect-assisted braiding procedure where the zero mode is not fully localized before and after passing through the local defect, but retains a residual coupling $\delta$ with a portion of the tri-junction, as shown in Fig.~\ref{fig:PassProcess}. We numerically compute the change in average fidelity as a function of the residual coupling strength $\delta$. The results, shown in Fig.~\ref{fig:defctst}, confirm that our scheme (with $\delta = 0$) effectively prevents errors.

We also calculate the unitarity (defined in Methods~\ref{sec:fidelity}) of the braiding matrix. As shown in Fig.~\ref{fig:defctst}, the unitarity follows a behavior similar to that of the average fidelity, confirming that errors arise when the local defect fails to fully return to its original state. This leads to components outside the subspace $\mathcal{H}_0$ in the states $\ket{\beta_1(T)}$ and $\ket{\beta_2(T)}$, resulting in a non-unitary braiding matrix. Setting $\delta=0$ in perfect defect-assisted braiding solves this problem. It also suggests that measuring the local defect state offers a practical method to assess braiding results.

\end{document}